\documentclass[11pt]{article}
\usepackage{amssymb}
\usepackage[all,dvips]{xypic} \CompileMatrices

\newcommand{\lbd}{\lambda}
\newcommand{\fun}{\rightarrow} 
\newcommand{\union}{\cup}
\newcommand{\appl}{\mbox{\raise.28ex\hbox{\tiny $\bullet$}}}
\newcommand{\set}[1]{\{#1\}}
\newcommand{\pair}[2]{\langle #1 , #2 \rangle}
\newcommand{\qqs}[2]{\forall\, #1\;\: #2}
\newcommand{\xst}[2]{\exists\, #1\;\: #2}
\newcommand{\ol}[1]{\overline{#1}}	
\newcommand{\ul}[1]{\underline{#1}}
\newcommand{\setc}[2]{\set{#1 \mid #2}}

\newcommand{\mor}{\mbox{ or }}

\newcommand{\pcf}{\mbox{\sc Pcf}}
\newcommand{\ff}{{\it ff}}
\newcommand{\Arrow}{\Rightarrow}

\newcommand{\hfs}[1]{\stackrel{#1}{\longrightarrow}} 
\newcommand{\state}[3]{(#1\; #2\; #3)}
\newcommand{\ppm}[2]{[#1,\stackrel{#2}{\hookleftarrow}]}   
\newcommand{\cppm}[2]{[#1,{\tt #2}]}   
\newcommand{\pom}[1]{#1}    
\newcommand{\mpom}[2]{\pair{#1}{{\bf #2}}}   
\newcommand{\dyn}[1]{{\bf #1}} 
\newcommand{\spom}[2]{\pair{#1}{#2}}  
\newcommand{\sppm}[2]{[#1,#2]} 
\newcommand{\comm}[2]{\stackrel{\fbox{\footnotesize $#1$}}{#2}}
\newcommand{\erase}[1]{{\it erase}(#1)}
\newcommand{\position}[2]{#1\appl #2}  
\newcommand{\pop}{{\it pop}}
\newcommand{\dnumber}[1]{\pi'(#1)}
\newcommand{\litem}[2]{\{#2\leftarrow #1\}}
\newcommand{\slitem}[3]{\{#2\stackrel{#3}{\leftarrow} #1\}}
\newcommand{\jitem}[2]{\pair{#1}{#2}} 
\newcommand{\vitem}[2]{\pair{#1}{#2}}
\newcommand{\jump}[3]{{\it jump}_{#1}^{#2}(#3)}
\newcommand{\view}[2]{{\it view}_{#1}(#2)} 
\newcommand{\dview}[2]{{\it dview}_{#1}(#2)} 
\newcommand{\case}{{\it case}\; }
\newcommand{\fcase}[2]{\mbox{{\it case}}\; #1\; [#2]}
\newcommand{\lcont}[2]{#1 \rightarrow #2}
\newcommand{\letcbv}[3]{{\it let}\: #1=#2\:{\it in}\:#3}

\newcommand{\dom}[1]{{\it dom}(#1)}
\newcommand{\hd}[1]{{\it hd}(#1)}
\newcommand{\Sub}[2]{[#1 \leftarrow #2]}
\newcommand{\cons}{\mbox{\raise.28ex\hbox{\tiny $\bullet$}}}
\newcommand{\placeh}{\mbox{\raise.28ex\hbox{\footnotesize $\bullet$}}}
\newcommand \seql[3]{\raisebox{3ex}{$\mbox{#1}\;$} \; \shortstack{$#2$ \\ \mbox{}\\
                    \mbox{}\hrulefill\mbox{}\\ \mbox{}\\ $#3$}}
\newcommand \seq[2]{\shortstack{$#1$ \\ \mbox{}\\
                    \mbox{}\hrulefill\mbox{}\\ \mbox{}\\ $#2$}}

\newcommand{\Figbar}{{\center \rule{\hsize}{0.5mm}}}  

\newcommand{\trad}[2]{\lceil #1 \rceil^{\it #2}} 
\newcommand{\samenv}[1]{\trad{\Gamma_{#1}}{GS}}   

\newcommand{\coerce}[1]{{\tt c}(#1)}  

\newcommand{\demon}{\maltese}

\newcommand{\coupe}[2]{\langle #1 \:|\: #2 \rangle}

\newcommand{\appst}[1]{\ol{#1}}   

\newenvironment{branch}{\left\{\begin{array}{l}}{\end{array}\right.}

\begin{document}

\title{Abstract Machines for Dialogue Games}
\author{P.-L. Curien (CNRS - Paris 7)\\ H. Herbelin (INRIA-Futurs)}
\maketitle

\begin{abstract}
The notion of abstract B\"ohm tree has arisen as an operationally-oriented distillation of
works on game semantics, and has been investigated in two papers \cite{Cur97, CuHer98}.
This paper revisits the notion, providing more syntactic support and more examples (like call-by-value evaluation) illustrating  the generality of the underlying computing device.  Precise correspondences between various formulations of the evaluation mechanism of abstract B\"ohm trees are established.
\end{abstract}

\section{Introduction}

This paper is a contribution to the paradigm of computation as interaction, by which we mean that 
a computation is described in terms of a game between two players, one representing the expression to be computed, the other its context, containing information such as the values of its free variables, or where the result should be returned to. This line of work has been pursued in different, but related perspectives, giving rise to rich theories and applications.

\begin{itemize}
\item The theory of sequential algorithms of Berry and Curien \cite{BeCu1,BeCu2} arose in the investigation of the full abstraction problem for PCF, a famous problem in the semantics of programming languages. PCF is a core pure functional programming language \cite{Plotkin77}, and a fully abstract model is a model capturing the observable differences between programs exactly. 
Sequential algorithms are mathematical objects that in addition to input-output behaviour record some information about the order of computation. They turned out to  provide a fully abstract model, not of PCF, but of a natural extension of PCF with a non-local control operator \cite{CarCuFel}.
\item Linear logic, and one of its models in particular -- the geometry of interaction -- originated in a fine-grained analysis of the cut-elimination process in proof theory, and has brought a wealth of new insights, such as formulas as resources, or proof nets \cite{Gir87,Gir89}.
\item Game semantics \cite{Coq95,HO,AJM} was triggered by these previous works, and has allowed to give a neat account of a variety of programming features, such as control, non-determinism, references...
\end{itemize}

In this paper, we adopt a type-free, operationally-oriented view. Our work takes inspiration mostly from the works of Coquand \cite{Coq95} and of the second author \cite{Her95,Her97}, and from those of Hyland, Ong, and Nickau \cite{HO,Nickau}.  Our key object is the notion of abstract B\"ohm tree, which is a generalization of that of B\"ohm tree. B\"ohm trees are (potentially infinite) normal forms, and play an important role in the theory of the $\lbd$-calculus \cite{Bare}. 
The main benefit of the generalization is that it offers the right level of generality for explaining the mechanism of computation at hand in the $\lbd$-calculus and similar sequential languages.
Abstract B\"ohm trees have been defined and studied in the two articles  \cite{Cur97, CuHer98}.
Here, we revisit the notion: we provide more syntactic support and more examples (like call-by-value evaluation) illustrating  the generality of the underlying computing device.  Precise statements on the correspondences between various formulations of the evaluation mechanism of abstract B\"ohm trees are established. 

\medskip
The paper is organized as follows.  Abstract B\"ohm trees are defined in section \ref{GAM-section}, where we also introduce our computational engine, called the Geometric Abstract Machine (GAM).
A concrete term notation with bound variables, in the style of the $\lbd$-calculus, is introduced at the end of this section. Section \ref{Examples-section} is devoted to examples, that cover $\lbd$-calculus (both normal and non-normal forms), and extensions: \pcf\ and classical \pcf;
call-by-value evaluation is also treated, and we show finally how Girard's ludics fits in our framework. A remarkable feature of our framework is that the computing device need not be extended or adjusted: only the compilation of the different source languages varies, and the machinery of abstract B\"ohm trees works  as a  ``universal" device.  

Sections \ref{VAM-section} and \ref{SAM-section} propose equivalent formulations of the GAM: the View Abstract Machine (VAM) highlights  the important notion of view (basic to the works  of Coquand \cite{Coq95}, and of Hyland and Ong \cite{HO}), while the Environment Abstract Machine is a straightforward generalization of (a stack-free version) of Krivine Abstract Machine \cite{Krivine-KAM}.
In the appendix, we establish precise correspondences between  these machines.

In section \ref{Strong-section}, we show how to formalize a lazy, stream-like computational loop calling the GAM again and again in order to produce the full result of a composition; each call of the GAM gets us to the (abstract B\"ohm tree version of the) next head variable of the composition along a given exploration path. In section \ref{non-normal}, we show how to extend the formalism of abstract B\"ohm trees and the GAM
to ``non-normal forms".

Finally, in section \ref{eta-section}, we discuss $\eta$-expansion, which is needed to evaluate (the compilation of) untyped $\lbd$-terms.
In this section, we also discuss the property of separation, which is the ability of observing differences through execution against a fixed counter-strategy.

\section{The Geometric Abstract Machine} \label{GAM-section}

In this section, we present the ingredients of our theory, starting with moves, positions, strategies and counter-strategies (section \ref{Pos-Strat}), and continuing with our computing device governing the interaction strategy / counter-strategy, the Geometrical Abstract Machine (section \ref{GAM-machine}). To this effect, we introduce the notions of multiplexed position, multiplexed strategy, multiplexed counter-strategy (section \ref{multiplexing}), which accommodate the process of duplication in the course of computation (when a function calls its argument several times). The termination cases of the machine are spelled out  (section 
\ref{termination}). A new contribution of this paper is section \ref{syntax-abt}, where we provide a term notation for abstract B\"ohm trees.

\subsection{Positions and strategies} \label{Pos-Strat}
We suppose given an 
alphabet $A$ of move names, containing a special symbol $\bullet$, which is
the initial move. Positions are sequences of moves with backward 
pointers for
player's moves (that is, moves occurring at even places in the position). 
We choose to 
represent pointers by numbers which count the number of opponent's moves 
between the pointing player's move and the pointed opponent's move. These 
pointers may be used to relate the bound occurrences to their binders, 
or to relate
values to their return address -- i.e., to the root of the subexpression of which
they are a (possible) value. Both of these kinds of pointing structure 
are present in the language \pcf\ (see section \ref{Examples-Pcf}).
An even position, or {\em response}, i.e., a position of even length, is a sequence of the 
form:
$$\pom{a_1}\ppm{a_2}{i_1}\ldots\pom{a_{2n-1}}\ppm{a_{2n}}{i_n}$$
where $a_j\in A$ for all $j\leq 2n$ and $i_l\in\omega\union\set{\_}$ 
for all $l\leq n$. An odd position, or {\em query},  is 
defined in the same way, but ends with an opponent's move $\pom{a_{2n-1}}$. 
In a position $p\ppm{a}{i}$, with $i\in\omega$, the intention is that the move
$\ppm{a}{i}$ points to the move $\pom{b}$ of $p$ which is at 
distance $2i+1$ from $\ppm{a}{i}$. 
For instance, if $i=0$ and $p=p_1\pom{b}$, then $\ppm{a}{i}$ points to
$\pom{b}$.
The number $i$ has to be small enough to guarantee that the corresponding 
$\pom{b}$ exists. We {\em will always assume this}, and it will be an (easy)  invariant of all the abstract machines presented in this paper that these pointers never become dangling while execution progresses.
We use $\_$ to designate free occurrences of player's moves. 

We shall let $q$ and $r$ range over queries and responses, respectively. We shall use
$\ppm{a}{\kappa}$ to designate either $\ppm{a}{i}$ or $\ppm{a}{\_}$.

\medskip
A {\em strategy} is a set $\phi$ of positions sucht that: 
\begin{itemize}
\item[-] all positions of $\phi$ are of even length,
and of the form $\bullet p$, where $\bullet$
does not occur in $p$,
\item[-] $\phi$ is closed under prefix,
\item[-] if $q\ppm{a_1}{\kappa_1},q\ppm{a_2}{\kappa_2}\in\phi$, then 
$\ppm{a_1}{\kappa_1}=\ppm{a_2}{\kappa_2}$.
\end{itemize}
The last property ensures that a ``query'', 
that is, 
a position of odd length, is uniquely answered in a strategy. 
Another presentation of a 
strategy is as a partial function, also written $\phi$, from queries
(whose player's moves are irrelevant) to player's moves . We write $\dom{\phi}$ to denote the domain 
of definition of this partial function. We shall freely use either of the two presentations.

A counter-strategy is a forest of strategies, where the roots are renamed so as to hook-up with the free moves of the strategy against which they are placed to play with. The renaming is defined as follows: 
$$\Sub{a}{\phi}=\setc{ar}{\bullet r\in \phi}$$
Hence  $\Sub{a}{\phi}$'s root is labelled by $a$.
A counter-strategy $\psi$ is a  union of renamed 
strategies $\Sub{a_1}{\phi_1},\ldots,\Sub{a_n}{\phi_n}$ (with all the 
$a_i$'s distinct and $\neq\bullet$). We write $\psi$  as:
$$\psi=[a_1\leftarrow\phi_1,\ldots,a_n\leftarrow \phi_n]\;$$
We assume that $\bullet$ is not only  the initial move of all positions of
 $\phi$, but does not occur either in $\psi$. (These conventions about $\bullet$ apply everywhere in the paper except in section \ref{Examples-CBV}, where $\bullet$ will be a ``real"  move expressing the convergence of a function in the weak sense, i.e. the presence of a head $\lbd$.)

\smallskip
Nothing prevents us from having infinite horizontal branching after player's moves, although in most examples
branching will only be finite.  Nothing prevents us either from having infinite positions and infinite depth strategies.

\subsection{Multiplexing}  \label{multiplexing}
Next we introduce multiplexed strategies, which will serve to trace dialogues 
between strategies 
and counter-strategies. The idea is 
that during the course of evaluation, nodes may be visited several times, 
whence the idea of ``opening new copies''  (see also section
\ref{VAM-discussion}).
A {\em multiplexed} even position is a sequence $\dyn{p}$ of the form:
$$\mpom{a_1}{j_1}\ppm{a_2}{i_1},\ldots,
\mpom{a_{2n-1}}{j_n}\ppm{a_{2n}}{i_n}$$
where the $a$'s and the $i$'s are as for positions, 
and where $j_1,\ldots,j_n\in\omega$ encode 
the multiplexing of opponent's moves. 
In \cite{CuHer98}, we used the terminology ``dynamic" for what we call ``multiplexed" here.
The new terminology reflects better the underlying idea of duplication. We could also use the word ``thick", following \cite{Boudes2005}.

A multiplexed strategy is defined as a tree $\Phi$ of multiplexed positions 
respecting the same 
conditions as a strategy, plus the following one: 
the collection of opponent's moves 
$\mpom{a}{j}$ occurring in $\Phi$ is in one-to-one correspondence with the 
set of their 
second components ${\bf j}$. In other words, the multiplexing indices describe 
a traversal of the multiplexed
tree. Moreover, if $\mpom{a}{j}$ appears in a multiplexed position 
$\dyn{p}$ of $\Phi$, then all the $\mpom{a'}{j'}$'s occurring before 
$\mpom{a}{j}$ in $p$ must be such that $j'<j$. 
This is a common constraint of tree traversals: 
a node cannot be visited unless all its descendants have been visited before.

\subsection{The machine} \label{GAM-machine}

\medskip
The Geometric Abstract Machine, or GAM, is a simple device describing 
the interaction between a strategy 
$\phi$ and a counter-strategy $\psi$. The machine duplicates progressively 
and in alternation
(greater and greater portions of) $\phi$ and $\psi$. 
The state of the machine consists of a sequence $\Gamma$ of multiplexed 
positions:
$$\begin{array}{lll}
\mbox{GAM states:} &&
\Gamma ::= \litem{\mpom{\bullet}{1}}{1} \mid \Gamma\litem{\dyn{p}}{\nu}
\end{array}$$
The successive items of $\Gamma$ are numbered 
$1,\appst{2},2,\appst{3},3,\ldots,n,\appst{n+1},\ldots$ (and we use $\nu$ to range over these step numbers).
The sequence $\Gamma$ can be put apart, yielding two multiplexed strategies:
\begin{eqnarray*}
\Phi & = & \setc{p}{\xst{n}{\position{\Gamma}{(2n-1)}=p}
\mor\position{\Gamma}{\appst{2n}}=p}\\
\Psi & = & \setc{p}{\xst{n}{\position{\Gamma}{(2n)}=p}
\mor\position{\Gamma}{\appst{2n+1}}=p}
\end{eqnarray*}
Clearly, $\Phi$ and $\Psi$ keep exactly the same information as $\Gamma$, thanks to the time stamps embodied in the opponent's moves.  And while it is simpler to write $\Gamma$ than  to write the pair of $\Phi$ and $\Psi$, it is really $\Phi$ and $\Psi$ that we have in mind, and that we shall draw in examples.

\medskip
\noindent
Here is some additional notation.
\begin{itemize}
\item[-] Given an odd multiplexed position $\dyn{q}$ ending with $\mpom{a}{n}$, we set 
$n=\dnumber{\dyn{q}}$.
\item[-] Given a multiplexed position $\dyn{p}$, $\erase{\dyn{p}}$ is the position obtained 
by erasing the
multiplexing information from $\dyn{p}$.
\item[-] Given an odd multiplexed position $\dyn{q}=\dyn{q}_1\ppm{a_1}{i}\mpom{a_2}{j}$, 
we write $\dyn{q}_1=\pop(\dyn{q})$. 
\end{itemize}

The transition rule of the GAM are presented in figure
\ref{GAM-rules}. Notice that the transitions $(\appst{2n})$ and $(\appst{2n+1})$ are essentially the same, 
the only difference being the exchange 
of $\phi$ and $\psi$. Similarly, the only difference between the transitions
$(2n)_b$ and $(2n+1)$ lies in parities. The transition $(2n)_f$ is a variation
of $(2n)_b$ which is linked to our choice of encoding for free variables.

\begin{figure}
\Figbar
$$\seql{(1)}{}{\mapsto\litem{\mpom{\bullet}{1}}{1}}$$

$$\seql{$(\appst{2n})$}{\hd{\Gamma}=\litem{\dyn{q}}{2n-1}\quad
\phi(\erase{\dyn{q}})=\ppm{a}{\kappa}}
{\Gamma\mapsto\Gamma\litem{\dyn{q}\ppm{a}{\kappa}}{\appst{2n}}}$$

$$\seql{$(2n)_b$}{\hd{\Gamma}=\litem{\dyn{q}\ppm{a}{i}}{\appst{2n}}\quad
\dnumber{\pop^i(\dyn{q})}=2m-1\quad
 \position{\Gamma}{\appst{2m-1}}=\dyn{r'}}
{\Gamma\mapsto\Gamma\litem{\dyn{r'}\mpom{a}{2n}}{2n}}$$

$$\seql{$(2n)_f$}{\hd{\Gamma}=\litem{\dyn{q}\ppm{a}{\_}}{\appst{2n}}}
{\Gamma\mapsto\Gamma\litem{\mpom{a}{2n}}{2n}}$$

$$\seql{$(\appst{2n+1})$}{\hd{\Gamma}=\litem{\dyn{q}}{2n}\quad
\psi(\erase{\dyn{q}})=\ppm{a}{\kappa}}
{\Gamma\mapsto\Gamma\litem{\dyn{q}\ppm{a}{\kappa}}{\appst{2n+1}}}$$

$$\seql{$(2n+1)$}{\hd{\Gamma}=\litem{\dyn{q}\ppm{a}{i}}{\appst{2n+1}}\quad
\dnumber{\pop^i(\dyn{q})}=2m\quad
\position{\Gamma}{\appst{2m}}=\dyn{r'}}
{\Gamma\mapsto\Gamma\litem{\dyn{r'}\mpom{a}{2n+1}}{2n+1}}$$
\caption{The Geometric Abstract Machine (GAM)}
\label{GAM-rules}
\Figbar
\end{figure}

Less formally, the transitions $(\appst{2n})$ and $(2n)_b$ 
can be described as follows:
\begin{itemize}
\item[-] 
At stage $(2n-1)$, the machine has reached (a copy of)  an opponent's position
in $\phi$. It looks 
up in $\phi$ a uniquely determined player's move 
$\ppm{a}{\kappa}$, and feeds it 
in $\Gamma$. 
\item[-] The machine is now at stage $(\appst{2n})$, and points to a position 
$\dyn{q}\ppm{a}{i}$. The machine uses the information $i$ to retrieve the stage 
$(2m-1)$ at which the prefix of $\dyn{q}$ to which $\ppm{a}{i}$ points has been 
built. At the stage $(\appst{2m-1})$ immediately preceding stage $(2m-1)$,
the machine pointed to a position $\dyn{r}'$ of $\psi$. Then the machine will 
place the move $a$ right at the end of $\dyn{r}'$, together with a multiplexing 
information, which is conveniently encoded as the current machine stage $(2n)$.
\end{itemize}
In summary, two mechanisms are mixed together:
%





\begin{enumerate}
\item The determinacy of $\phi$ and $\psi$ are put to profit for 
resolving in turn conflicts about which among the possible opponent's moves pending
from a given player's position should be played next. 
\item The pointer structure together with the multiplexing structure is used 
to
determine under which player's position the next opponent's move is to be 
placed.
\end{enumerate}
We refer to these two ingredients of the machinery as to the tree interaction
and pointer interaction, respectively. 

\subsection{Termination} \label{termination}

There are exactly three situations in which the GAM is prevented to proceed further.
\begin{enumerate}
\item
Suppose that when attempting to perform step $(2n)_b$ we find $m=1$.
Then since there is no step $(\appst{1})$ the machine stops.
This is what happens in the example of section \ref{Examples-Pcf}: the final move points 
to the root of $\phi$. 
\item
Suppose that performing step $(\appst{2n+1})$ results in adding a move 
$\ppm{a}{\_}$. Then the machine is unable to perform step $(2n+1)$, because there is no counterpart of rule $(2n)_f$ for the counter-strategy.
This is what happens in the example of section 
\ref{Examples-Boehm}: the computation terminates because a free variable of $\psi$ 
has been met. 
\item
Let $\dyn{p}$ be the opponent's multiplexed position reached  at a stage $(2n-1)$, then
step $\appst{2n}$ can be performed only if $\erase{\dyn{p}}\in\dom{\phi}$
(and similarly for $(2n)$ and $\psi$).
 \end{enumerate}
 
 We make some observations:
 \begin{itemize}
 \item
In \cite{Cur97} we have defined a notion of (abstract) typing that guarantees that case (3) of termination never occurs. An important instance is that of typed, $\eta$-long B\"ohm  trees, where each occurrence of a variable is applied to all its arguments, and in each 
sequence of abstractions
$\lbd\vec{x}.M$ the number of parameters $x_1,\ldots,x_n$ is exactly 
the number of arguments $N_1,\ldots,N_n$ that the term $\lbd\vec{x}.M$ accepts, according to the types. 
We consider that case (3) of termination is improper, and
in section \ref{eta-section} we examine how to continue the execution when this case shows up.

\item
In \cite{Cur97} it is also proved, as a corollary of a result of Coquand \cite{Coq95}, that the GAM stops after a finite number of steps, provided both $\phi$ and $\psi$ have finite depth. This result holds without any typing assumptions, unlike usual termination results. This is because $\beta$-reduction is in fact
more liberal than the GAM and embodies implicit $\eta$-expansions (see section \ref{eta-section}), which are the only source of non-termination.

\item
Even when the machine stops satisfactorily (cases (1) or (2)), we may not have the final word on the
composition of $\phi$ and $\psi$.
Consider for example $(\lbd y.y(x))\Sub{x}{{\it tt}}$ in \pcf. Then the
GAM does not produce $(\lbd y.y({\it tt}))$, but $(\lbd y.y(x\Sub{x}{{\it tt}}))$. This is
similar to the situation with environment
machines, which do not compute under $\lbd$'s. In section \ref{Strong-section}, we show how to compute (arbitrary long positions) of the composition $\phi\circ\psi$.
\end{itemize}
\subsection{Syntax for abstract B\"ohm trees} \label{syntax-abt}

Rather than defining strategies as sets of positions, one can define them (co-recursively) trough the following equations, which specify an abstract syntax for abstract B\"ohm trees.

\smallskip
\noindent Abstract syntax:
$$\begin{array}{l}
P::= \ppm{a}{\kappa}\:\setc{(b,M_b)}{b\in B} \quad(\kappa=i\mbox{ or }\kappa=\_)\\
M::=(P)
\end{array}$$
Note that the root of a term $M$ is not named, unlike internal opponent's nodes ($M_b$ is named by $b$).
This is the syntactic counterpart of our convention to denote the initial move of a strategy by  the special symbol $\bullet$.

For example, the abstract term denoting 
$$\bullet\ppm{a_1}{\_}\begin{branch}
a_2\ppm{a_3}{1}\\
a_4\ppm{a_5}{0}
\end{branch}$$
is $(\ppm{a_1}{\_}\:\set{(a_2,(\ppm{a_3}{1}\set{\:})),(a_4,(\ppm{a_5}{0}\set{\:}))})$.

\medskip
This syntax is reminiscent of De Bruijn notation for the $\lbd$-calculus \cite{deBruijn72}, and hence by ``reverse engineering" suggests to replace pointers by bound variables. This leads us to the following concrete syntax:

\smallskip
\noindent Concrete syntax:
$$\begin{array}{l}
P::= \cppm{a}{\star}\:\setc{(b,M_b)}{b\in B}  \quad(\star={\tt x}\mbox{ or } \star=\_)\\
M::=(\lbd {\tt x}.P)
\end{array}$$

 \medskip\noindent
 There are free variables $\cppm{a}{\_}$, and bound variables $\cppm{a}{{\tt x}}$, where ${\tt x}$
stands for a set of variables bound at the same place. When ${\tt x}$ does not occur free in $P$,
we shall freely write just $(P)$ instead of $(\lbd {\tt x}.P)$ (note that the parenthesis is meaningful, as it signals  the player's move from $(P)$ to $P$).

\smallskip
 The compilation  from concrete to abstract syntax  (see section \ref{EAM-machine} for a translation in 
 the converse direction) takes as paramater a list of variable names (taken to be empty initially):
 
 $$\begin{array}{l}
 \trad{(\lbd {\tt x}.P)}{ca}_L=(\trad{P}{ca}_{{\tt x}\cons L})\\
 \trad{\cppm{a}{\_}\:\setc{(b,M_b)}{b\in B}}{ca}_L=\ppm{a}{\_}\:\setc{(b,\trad{M_b}{ca}_L)}{b\in B}\\
  \trad{\cppm{a}{{\tt x}}\:\setc{(b,M_b)}{b\in B}{ca}_L}=\ppm{a}{L_{{\tt x}}}\:\setc{(b,\trad{M_b}{ca}_L)}{b\in B}\
 \end{array}$$
 where 
 $$\begin{array}{lll}
 \seq{}{({\tt x}\cons L)_{{\tt x}}=0} &&
  \seq{L_{{\tt x}}=i\;\; {\tt x}\neq{\tt y}}{({\tt y}\cons L)_{{\tt x}}=i+1}
 \end{array}$$
 
 \noindent
 The Geometrical Abstract Machine can be formulated in terms of this syntax (see section \ref{EAM-machine}).

\section{Examples} \label{Examples-section}

In this section, we present a collection of examples of abstract B\"ohm trees. We start with the simplest and motivating example of B\"ohm trees (or $\lbd$-terms in normal form)  (section \ref{Examples-Boehm}). (We will show later how to fit $\lbd$-terms in general (section \ref{non-normal}), up to the problem of $\eta$-conversion, which is discussed separately (section \ref{eta-section}).) 
We next move on  to a small variation, the \pcf\ trees, which are $\lbd$-terms with case statements (section \ref{Examples-Pcf}). A further variation is to add control, \`a la $\lbd\mu$-calculus \cite{Parigot92} (section \ref{Examples-Classical}). The last two examples are new to this paper: in section \ref{Examples-CBV},
we present {\em call-by-value} trees and show how to compile them into abstract B\"ohm trees, and in section \ref{Examples-Ludics} we show how to fit Girard's designs (the notion central to ludics \cite{GirLS}) in our framework. These examples reinforce the generality and relevance of the notion of abstract B\"ohm tree. It is in particular remarkable that in order to execute call-by-value we do not need to change our engine: all happens during the compilation, much like in continuation-passing-style translations \cite{Plotkin75}.

\subsection{B\"ohm trees} \label{Examples-Boehm}

We represent 
B\"ohm trees, or (potentially infinite) $\lbd$-terms in normal form, as trees with pointers. The traditional
nodes $\lbd x_1\ldots x_m.y$ of B\"ohm trees are split in a pair of 
moves: an opponent's move $\lbd\vec{x}$ and a player's move $y$.
$$\begin{array}{l}
B::= (\lbd x_1\ldots x_m.P)\\
P::= yB_1\ldots B_n
\end{array}$$
The alphabet is $A=\omega\cup{\it Var}$, where the variables
$x\in{\it Var}$
are only used as player's moves of the form $\ppm{x}{\_}$. 

\smallskip
We next show how to compile B\"ohm trees (through the concrete syntax of section \ref{syntax-abt}).
For defining the translation, it is convenient to prepare the source term in such a way that each bound variable has an indexed format ${\tt x}_i$, and that each free variable has a non-indexed format $x$.
Then the translation into the concrete syntax of abstract B\"ohm trees is straightforward:

$$\begin{array}{l}
\trad{xB_1\ldots B_n}{}=\cppm{x}{\_}\set{(1,\trad{B_1}{}),\ldots,(n,\trad{B_n}{})}\\
\trad{{\tt x}_iB_1\ldots B_n}{}=\cppm{i}{{\tt x}}\set{(1,\trad{B_1}{}),\ldots,(n,\trad{B_n}{})}
\\\\
\trad{\lbd {\tt x}_1\ldots{\tt x}_m.P}{}=(\lbd{\tt x}.\trad{P}{}) \quad (m\geq 0)
\end{array}$$

We illustrate this with an example. 

\medskip

Strategy for $(u(\lbd x. u(\lbd y.x)))$:

$$\comm{(}{\pom{\bullet}}\comm{u}{\ppm{u}{\_}}\begin{branch}
\comm{(\lbd x}{\pom{1}}\comm{u}{\ppm{u}{\_}}\begin{branch}
\comm{(\lbd y}{\pom{1}}\comm{x}{\ppm{1}{1}}\end{branch}
\end{branch}$$

Strategy for $\Sub{u}{(\lbd r. r(r(z)))}$
$$\comm{(\lbd r}{\pom{u}}\comm{r}{\ppm{1}{0}}\begin{branch}
\comm{(}{\pom{1}}\comm{r}{\ppm{1}{1}}\begin{branch}
\comm{(}{\pom{1}}\ppm{z}{\_}\end{branch}
\end{branch}$$

The trace of the execution of 
$(u(\lbd x. u(\lbd y.x))\Sub{u}{(\lbd r. r(r(z)))}$, 
is displayed below.  The boxed information (in this example, and in others to follow) maintains some 
 reminders of the underlying concrete syntax that may  help the reader to follow what is going on.

\medskip

Function multiplexed tree:

$$\comm{(}{\mpom{\bullet}{1}}\comm{u}{\ppm{u}{\_}}\begin{branch}
\comm{(\lbd x}{\mpom{1}{3}}\comm{u}
{\ppm{u}{\_}}\begin{branch}
\comm{(\lbd y}{\mpom{1}{5}}\comm{x}{\ppm{1}{1}}\end{branch}\\
\comm{(\lbd x}{\mpom{1}{7}}\comm{u}{\ppm{u}{\_}}\begin{branch}
\comm{(\lbd y}{\mpom{1}{9}}\comm{x}{\ppm{1}{1}}\end{branch}
\end{branch}$$

\medskip

Argument multiplexed tree:

$$\begin{branch}
\comm{(\lbd r}{\mpom{u}{2}}\comm{r}{\ppm{1}{0}}\begin{branch}
\comm{(}{\mpom{1}{6}}\comm{r}{\ppm{1}{1}}\begin{branch}
\comm{(}{\mpom{1}{10}}\ppm{z}{\_}\end{branch}
\end{branch}\\
\comm{(\lbd r}{\mpom{u}{4}}\comm{r}{\ppm{1}{0}}\\
\comm{(\lbd r}{\mpom{u}{8}}\comm{r}{\ppm{1}{0}}
\end{branch}$$

Here the strategy and counter-strategy are just paths, hence we could hardly
illustrate the tree interaction. But we did illustrate pointer interaction. 
Let us describe the first steps of the execution. We start by applying rule $(1)$, which lets us point to the root of the strategy:

\medskip
$\comm{(}{\mpom{\bullet}{1}}$

\medskip\noindent
Then we apply rule 
 $(\appst{2})$, which leads us to a free move:
 
 \medskip
 $\comm{(}{\mpom{\bullet}{1}}\comm{u}{\ppm{u}{\_}}$
 
 \medskip\noindent
Therefore, we apply rule $(2)_f$, and we point now to the root of the (unique tree of the) counter-strategy:

\medskip
$\comm{(\lbd r}{\mpom{u}{2}}$

\medskip\noindent
By rule $(\appst{3})$ we reach a move 1 which points to the move reached at step (2), let us say for short that it points to 2:
 
 \medskip
 $\comm{(\lbd r}{\mpom{u}{2}}\comm{r}{\ppm{1}{0}}$

\medskip\noindent 
Hence the move at step $(3)$ is played under $\appst{2}$:

\medskip
$\comm{(}{\mpom{\bullet}{1}}\comm{u}{\ppm{u}{\_}}\begin{branch}
\comm{(\lbd x}{\mpom{1}{3}}\end{branch}$

\medskip\noindent
Step $(\appst{4})$ leads us then to open a new copy of the counter-strategy: 

\medskip
$\comm{(}{\mpom{\bullet}{1}}\comm{u}{\ppm{u}{\_}}\begin{branch}
\comm{(\lbd x}{\mpom{1}{3}}\comm{u}
{\ppm{u}{\_}}\end{branch}$

\medskip\noindent
This is where multiplexing begins:

\medskip
$\comm{(\lbd r}{\mpom{u}{4}}$

\medskip\noindent
A little later, say, when we have performed step $(\appst{7})$, we arrive to a move 1 which points to 2:

\medskip
$\begin{branch}
\comm{(\lbd r}{\mpom{u}{2}}\comm{r}{\ppm{1}{0}}\begin{branch}
\comm{(}{\mpom{1}{6}}\comm{r}{\ppm{1}{1}}
\end{branch}\\
\comm{(\lbd r}{\mpom{u}{4}}\comm{r}{\ppm{1}{0}}\end{branch}$

\medskip\noindent
Then $(7)$ tells us to play $7$ under $\appst{2}$, opening a new copy of (a subtree of the) strategy:

\medskip
$\comm{(}{\mpom{\bullet}{1}}\comm{u}{\ppm{u}{\_}}\begin{branch}
\comm{(\lbd x}{\mpom{1}{3}}\comm{u}
{\ppm{u}{\_}}\begin{branch}
\comm{(\lbd y}{\mpom{1}{5}}\comm{x}{\ppm{1}{1}}\end{branch}\\
\comm{(\lbd x}{\mpom{1}{7}}
\end{branch}$

\medskip\noindent
Etc... (in general, when $\appst{n}$ points to $m$, then $n$ is to be placed under $\appst{m}$).
Note  the importance of keeping the multiplexing information
precise: for example, the move $\ol{10}$ points to $7$, not to $3$, and this
is precisely what gets us out of a loop.

\subsection{\pcf\ trees} \label{Examples-Pcf}

Our next example comes from the language \pcf.  We refer to \cite{AmaCur98} for background.
Let us just mention here that the trees presented here, which we call \pcf\  trees, provide a
term model for the language \pcf, a core functional programming language which has been
a subject of focus of many works in denotational semantics, much in the same way as
B\"ohm trees provide a term model for $\lbd$-calculus.
\pcf\ trees have the form 
$$\begin{array}{l}
B::=(\lbd \vec{x}.P)\\
P::=\case yB_1\cdots B_n\;[\ul{n_1}\Arrow B'_1,\ldots,
\ul{n_i}\Arrow B'_i,\ldots]
\end{array}$$
where the $\ul{n_i}$'s are distinct natural numbers (or 
boolean values).
(We use underlining for integer constants to distinguish them from the moves $i$ used in the compilation of 
variables and arguments.)
We compile \pcf\ B\"ohm trees much like B\"ohm trees, adopting the same convention about variables
(cf. section \ref{Examples-Boehm}):

$$\begin{array}{l}
\trad{\case xB_1\cdots B_n\;[\ldots,
\ul{n_i}\Arrow B'_i,\ldots]}{}=\cppm{x}{\_}\set{(1,\trad{B_1}{}),\ldots,(n,\trad{B_n}{}),\ldots,(\ul{n_i},\trad{B'_i}{}),\ldots}\\
\trad{\case {\tt x}_jB_1\cdots B_n\;[\ldots,
\ul{n_i}\Arrow B'_i,\ldots]}{}=\cppm{j}{{\tt x}}\set{(1,\trad{B_1}{}),\ldots,(n,\trad{B_n}{}),\ldots,(\ul{n_i},\trad{B'_i}{}),\ldots}
\\\\
\trad{\lbd {\tt x}_1\ldots{\tt x}_m.P}{}=(\lbd{\tt x}.\trad{P}{}) \quad (m\geq 0)
\end{array}$$

With respect to $\lbd$-calculus, we add the constants to the alphabet (in this section as well as in section
\ref{Examples-Classical}).
So, if  we limit ourselves to the type hierarchy built over ${\it Bool}$,
the alphabet is $A=\omega\cup{\it Var}\cup\set{{\it tt},\ff}$. The example below
illustrates 
tree interaction.

\medskip

Strategy for $(\case f ({\it tt}) \;[\ff\Arrow {\it tt}])$:

$$\comm{(}{\pom{\bullet}}\comm{\case f}{\ppm{f}{\_}}\begin{branch}
\comm{(}{\pom{1}}\ppm{{\it tt}}{0}\\
\pom{\ff}\ppm{{\it tt}}{1}
\end{branch}$$

\medskip

Strategy for $\Sub{f}{(\lbd x.\case x  \;[{\it tt}\Arrow \ff,\ff\Arrow {\it tt}])}$:

$$\comm{(\lbd x}{\pom{f}}\comm{\case x}{\ppm{1}{0}}\begin{branch}
\pom{{\it tt}}\ppm{\ff}{1}\\
\pom{\ff}\ppm{{\it tt}}{1}
\end{branch}$$

\medskip

Function multiplexed tree:

$$\comm{(}{\mpom{\bullet}{1}}\comm{\case f}{\ppm{f}{\_}}\begin{branch}
\comm{(}{\mpom{1}{3}}\ppm{{\it tt}}{0}\\
\mpom{\ff}{5}\ppm{{\it tt}}{1}
\end{branch}$$

\medskip

Argument multiplexed tree:

$$\comm{(\lbd x}{\mpom{f}{2}}\comm{\case x}{\ppm{1}{0}}\begin{branch}
\mpom{{\it tt}}{4}\ppm{\ff}{1}\\
\end{branch}$$

\subsection{Classical \pcf} \label{Examples-Classical}
The following variant of \pcf\ allows us to introduce explicit control on where the values are to 
be sent to.

$$\begin{array}{l}
M::=(\lbd\vec{x}.\mu\beta.c)\\
c::= \case x\vec{M}\:[\vec{\ul{n}}\rightarrow\vec{c}]\mid [\alpha]\ul{n}
\end{array}$$

\noindent
Terms of this syntax are called classical \pcf\ trees. Their execution is driven  by the following abstract machine:

$$\begin{array}{c}
\seq{\rho(x)=(\lbd\vec{z}\mu\beta.c')[\rho']}
{\coupe{\case x\vec{M}\:[\vec{\ul{n}}\rightarrow\vec{c}]}{\rho}  \quad\longrightarrow\quad 
\coupe{c'}{\rho'[\vec{z}\leftarrow\vec{M}[\rho],\beta\leftarrow[\vec{\ul{n}}\rightarrow\vec{c}][\rho]]}}
\\\\
\seq{\rho(\alpha)=[\vec{\ul{n}}\rightarrow\vec{c}[\rho']\mbox{ and } n=n_i}
{\coupe{[\alpha]\ul{n}}{\rho} \quad\longrightarrow\quad \coupe{c_i}{\rho'}}
\end{array}$$

\medskip\noindent
The compilation in concrete syntax is as follows (again, we assume that bound variables take an indexed format ${\tt x}_i$ and that free variables take a non-indexed format $x$):

$$\begin{array}{l}
\trad{(\lbd\vec{{\tt x}}\mu\beta.c)}{}_\rho=( \lbd {\tt x}.\trad{c}{}_{\rho[\beta\leftarrow x]})\\\\
\trad{[\alpha]\ul{n}}{}_\rho = (!\ul{n},\rho(\alpha))\\\\
\trad{\case xM_1\ldots M_m[\ldots,\ul{n_j}\rightarrow c_j,\ldots]}{}_\rho\\
\quad=\cppm{x}{\_}\:
\set{\ldots,(?i,\trad{M_i}{}_\rho),\ldots,(!\ul{n_j},(\trad{c_j}{}_\rho),\dots}\\\\
\trad{\case {\tt x}_iM_1\ldots M_m[\ldots,\ul{n_j}\rightarrow c_j,\ldots]}{}_\rho\\
\quad=\cppm{i}{{\tt x}}\:
\set{\ldots,(?i,\trad{M_i}{}_\rho),\ldots,(!\ul{n_j},(\trad{c_j}{}_\rho),\dots}
\end{array}$$

\subsection{Classical call-by-value \pcf} \label{Examples-CBV}

In call-by-value $\lbd$-calculus \cite{Plotkin75}, the $\beta$-rule $(\lbd x.M)N\rightarrow
\Sub{M}{x}{N}$ is allowed only when $N$ is a value, where a value $V$ is an abstraction $\lbd x.N'$ or a variable. Hence a non-value is a term of the form $MN$. Then, if $V,M_1$ are normal forms,  $(\lbd y.M_1)(xV)$ is also a normal form. It is more readable and suggestive to write the latter 
$\letcbv{y}{xV}{M_1}$.  Adding case statements and control as in section
\ref{Examples-Classical}, we arrive at the following syntax of {\em classical call-by-value B\"ohm trees}:

$$\begin{array}{ll}
\mbox{Values} &
V::=\lbd (z,\beta).c \mid \ul{n} \mid x \\
\mbox{Commands} & c::= \letcbv{x}{yV}{c} \mid
\case x\:[\vec{\ul{n}}\rightarrow\vec{c}]\mid [\alpha]V
\end{array}$$

\noindent
(An alternative syntax for ($\lbd(z,\beta)$ is $\lbd z\mu\beta$.) The expressions of the second category are called commands, and in a typed setting have the special type $\bot$ (see below).
In this syntax, a value $x$ is meant
to be of basic type (and is indeed of basic type in the typed version given at the end of the section).
This is not restrictive in terms of expressive power, as we can express higher-type variables $y$ as values in $\eta$-expanded form. For example, at first-order:
$$\lbd(z,\alpha).(\letcbv{u}{yz}{[\alpha]u})\;.$$
It is not restrictive either to limit application of a variable to one argument, as we can encode
$({\it let}\:xV_1V_2)$ as
$(\letcbv{z}{xV_1}{{\it let}\: y=g})$. Moreover, a key difference between call-by-name and call-by-value is that in call-by-value an abstraction is a value, or a result of computation. Hence the evaluation stops at the first abstraction met, while in call by--name, the machine treats abstractions by blocks and
proceeds until it finds the head variable. Compare for example
$$\begin{array}{lll}
\lbd xy.xM && \mbox{call-by-name} \\
\lbd(x,\alpha).[\alpha](\lbd(y,\beta).(\letcbv{z}{xV}{[\alpha] z}) && \mbox{call-by-value}
\end{array}$$
In the call-by-name case, the dialogue behind $\lbd xy.xM$ can be paraphrased as:
\begin{itemize}
\item[-] Question: What is the value of the head variable?
\item[-] Answer: $x$.
\end{itemize}
In the call-by-value setting, we must also know $y$, even it if does not occur in $V$. So the dialogue 
that we want to formalize is different:
\begin{itemize}
\item[-] Answer: I am an abstraction ($\lbd(x,\alpha)$).
\item[-] Question: If I give you an abstraction as value for $x$, what more can you tell me about you?
\item[-] Answer: I  shall become an abstraction ($\lbd(y,\beta)$) and my result should be returned to where $\alpha$ was declared bound.
\item[-] If I give you a value for $y$, what will you do next?
\item[-] I shall apply $x$ to $V$.
\end{itemize}
We hope that these intuitions will be helpful for the rest of this section, which is a bit technical.
The two syntactic categories are called values and commands, respectively.

\medskip The evaluation of our call-by-value trees is performed by the following abstract machine  (we refer to  \cite{CuHerDual}) for background on the notation $\tilde{\mu}$):

$$\begin{array}{l}
\seq{\rho(y)=(\lbd(z,\beta).c')[\rho']}
{\coupe{\letcbv{x}{yV}{c}}{\rho}\quad\longrightarrow\quad \coupe{c'}{\rho'\:[z\leftarrow V[\rho],\beta\leftarrow(\tilde{\mu}x.c)[\rho]]}}\\\\
\seq{\rho(x)=\ul{n}_i[\rho']}
{\coupe{\case x\:[\vec{\ul{n}}\rightarrow\vec{c}]}{\rho}  \quad\longrightarrow\quad 
\coupe{c_i}{\rho}}\\\\
\seq{\rho(\alpha)=(\tilde{\mu}x.c')[\rho']}
{\coupe{[\alpha]V}{\rho} \quad\longrightarrow\quad \coupe{c'}{\rho'}{x\leftarrow V[\rho]]}}
\end{array}$$

\medskip\noindent
Here is the compilation into our concrete syntax (we have been guided by the game semantics of call-by-value proposed by Honda and Yoshida \cite{Honda-value}):

$$\begin{array}{l}
\trad{\lbd (z^{\iota},\beta).c}{}_\rho
= \bullet\;\set{
\ldots,(?\ul{n},(\lbd{\tt x}.\trad{c}{}_{\rho[z\leftarrow \ul{n},\beta\leftarrow{\tt x}]})),\ldots} \quad({\tt x}\mbox{ fresh})\\
\trad{\lbd (z^{\sigma_1\rightarrow\sigma},\beta).c}{}_\rho
= \bullet\;\set{
(?\bullet,(\lbd{\tt x}.\trad{c}{}_{\rho[z\leftarrow{\tt x}, \beta\leftarrow{\tt x}]}))} \quad({\tt x}\mbox{ fresh})\\
\trad{\ul{n}}{}_\rho=\ul{n}\\
\trad{x^\iota}{}_\rho=\rho(x)\\\\
\trad{\case x\:[\vec{\ul{n}}\rightarrow\vec{c}]}{}_\rho=\trad{c_i}{}_\rho\quad(\rho(x)=\ul{n_i})\\
\trad{[\alpha]V}{}_\rho=\cppm{!*}{\rho(\alpha)}\;\set{\vec{T}}\quad(\rho(\alpha)\downarrow,\trad{V}{}_\rho=*\;\set{\vec{T}})\\
\trad{[\alpha]V}{}_\rho=\cppm{(!*,\alpha)}{\_}\;\set{\vec{T}}\quad(\rho(\alpha)\uparrow,\trad{V}{}_\rho=*\;\set{\vec{T}})\\
\trad{\letcbv{x^\iota}{yV}{c}}{}_\rho=\cppm{?*}{\rho(y)}\;\set{\vec{T},\ldots,(!\ul{n},(\trad{c}{}_{\rho[x\leftarrow \ul{n}]})),\ldots}
\quad(\rho(y)\downarrow,\trad{V}{}_\rho=*\;\set{\vec{T}})\\
\trad{\letcbv{x^\iota}{yV}{c}}{}_\rho=\cppm{(?*,y)}{\_}\;\set{\vec{T},\ldots,(!\ul{n},(\trad{c}{}_{\rho[x\leftarrow \ul{n}]})),\ldots}
\quad(\rho(y)\uparrow,\trad{V}{}_\rho=*\;\set{\vec{T}})\\
\trad{\letcbv{x^{\sigma_1\rightarrow\sigma}}{yV}{c}}{}_\rho=\cppm{?*}{\rho(y)}\;\set{\vec{T},(!\bullet,(\lbd{\tt z}.\trad{c}{}_\rho[x\leftarrow{\tt z}))}
\quad(\rho(y)\downarrow,\trad{V}{}_\rho=*\;\set{\vec{T}}, {\tt z}\mbox{ fresh})\\
\trad{\letcbv{x^{\sigma_1\rightarrow\sigma}}{yV}{c}}{}_\rho=\cppm{(?*,y)}{\_}\;\set{\vec{T},(!\bullet,(\lbd{\tt z}.\trad{c}{}_\rho[x\leftarrow{\tt z}]))}
\quad(\rho(y)\uparrow,\trad{V}{}_\rho=*\;\set{\vec{T}},{\tt z}\mbox{ fresh})\\\\
\trad{c\;[\ldots,x_i\leftarrow \ul{n_i},\ldots, x_j\leftarrow V_j^{\iota\rightarrow\sigma},\ldots,x_k\leftarrow V_k^{(\sigma_3\rightarrow\sigma_2)\rightarrow\sigma_1},\ldots]}{}_{[]} =\\
\quad \trad{c}{}_{[\ldots x_i\leftarrow \ul{n_i},\ldots]} \;[\ldots,(?\ul{n},x_j)\leftarrow(?\ul{n},T_n),\ldots,
(?\bullet,x_k)\leftarrow(?\bullet,T)]\\
\quad\quad
(\trad{V_j}{}{}_{[]}=\bullet\:\set{\ldots,(?\ul{n},T_n),\ldots},
\trad{V_k}{}_{[]}=\bullet\:\set{(?\bullet,T)})
\end{array}$$

The following explanations should help parsing the definition. The compilation is relative to an environment, which records the values of variables of basic types, and which also rearranges the
names of bound variables so as to pass correctly from the syntax of call-by-value B\"ohm trees to the syntax of abstract B\"ohm trees. The compilation of $\lbd(z,\beta).c$ is conform to the informal dialogue suggested above. The compilations of $\ul{n}$ and $x$ are self-explanatory (remember that $x$ must have a basic type in our syntax). As for commands, the compilation of $\case x\:[\vec{\ul{n}}\rightarrow\vec{c}]$ is easy, since the environment $\rho$ has ``precomputed" the branch to be chosen. The compilation of $[\alpha]V$ splits into two cases
according to whether $\alpha$ is bound or free. The compilation of $\letcbv{x}{yV}{c}$ splits into four cases, taking additionally into account the type (basic or not) of $x$. The common idea between the four cases for the ${\it let}$ construct is that the compilation of $\letcbv{x}{yV}{c}$ with respect to $\rho$ is obtained by adding sons to the root of the compilation of $V$ (with respect to $\rho$) corresponding to the different possible values of $yV$, under which the compilation of $c$ (with respect to $\rho$ extended with a suitable value for $x$) is placed. 

The notation $(?\ul{n},x_j)\leftarrow(?\ul{n},T_n)$  is an obvious variation on the notation $[a\leftarrow\phi]$ of section \ref{Pos-Strat}. (Note also that in this section $\bullet$ is a player's move from the point of view of the strategy $\phi$. In call-by-value we use the symbol $\bullet$ to denote the information ``I am a function".)
Finally, we notice that the translation  fixes some details about the names of moves. Our use of $!$ and $?$ here  is somewhat reminiscent of  the use of $n$ and $\ul{n}$  in the two previous sections to distinguish between arguments and continuations.

\smallskip
The alphabet in this section is thus the following:
$$A::= (?\bullet,\xi) \mid (?n,\xi) \mid ?\bullet \mid ?n \mid !\bullet \mid !n$$
where $\xi$ ranges over ordinary variables $x$ and continuation variables $\alpha$.

\medskip As an illustration, take:
$$\begin{array}{l}
\letcbv{y}
{x(\lbd (z,\beta).\fcase{z}{\lcont{3}{[\beta]5},\lcont{4}{[\beta]9}})}
{\letcbv{u}{vy}{[\alpha]u}}\\
\quad [x\leftarrow \lbd (t,\gamma).\letcbv{r}{t3}{\fcase{r}{\lcont{5}{[\gamma]7}}}\;,\;
v\leftarrow \lbd (w,\delta).\fcase{w}{\lcont{7}{[\delta]8}}]
\end{array}$$

\noindent
The compilation gives:
$$\begin{array}{l}
\ppm{(?\bullet,x)}{\_}\set{(?3,(\lbd{\tt x}_1.\cppm{!5}{{\tt x}_1})),
(?4,(\lbd{\tt x}_1.\cppm{!9}{{\tt x}_1})),..,\\
\quad\quad\quad\quad\quad(\pom{!a},(\ppm{(?a,v)}{\-}
\set{..,(\pom{!b},(\ppm{(!b,\alpha)}{\_})),..})),..}\\\\
\begin{branch}
(\pom{(?\bullet,x)},(\lbd{\tt x}_2.\ppm{?3}{{\tt x}_2}\set{(!5,(\ppm{!7}{{\tt x}_2}}))\\\\
(\pom{(?7,v)},(\lbd{\tt x}_3.\ppm{!8}{{\tt x}_3}))
\end{branch}
\end{array}$$

\noindent Or, as trees with pointers:

\smallskip
\begin{center}
\begin{tabular}{ll}
$\phi$ &
$\ppm{(?\bullet,x)}{\_}\begin{branch}
\pom{?3}\ppm{!5}{0}\\
\pom{?4}\ppm{!9}{0}\\
\vdots\\
\pom{!a}\ppm{(?a,v)}{\-}\begin{branch}
\vdots\\
\pom{!b}\ppm{(!b,\alpha)}{\_}\\
\vdots\end{branch}\\
\vdots
\end{branch}$\\\\
$\psi$ &
$\begin{branch}
\pom{(?\bullet,x)}\ppm{?3}{0}\pom{!5}\ppm{!7}{1}\\\\
\pom{(?7,v)}\ppm{!8}{0}
\end{branch}$
\end{tabular}
\end{center}

\noindent Here is the execution:

\begin{center}
\begin{tabular}{ll}
$\phi$ &
$\ppm{(?\bullet,x)}{\_}\begin{branch}
\mpom{?3}{3}\ppm{!5}{0}\\
\mpom{!7}{5}\ppm{(?7,v)}{\_}\begin{branch}
\mpom{!8}{7}\ppm{(!8,\alpha)}{\_}
\end{branch}\\
\end{branch}$\\\\
$\psi$ &
$\begin{branch}
\mpom{(?\bullet,x)}{2}\ppm{?3}{0}\mpom{!5}{4}\ppm{!7}{1}\\\\
\mpom{(?7,v)}{6}\ppm{!8}{0}
\end{branch}$
\end{tabular}
\end{center}

\noindent
Notice that the first move is $\appst{2}$: in call-by-value, it is Player who starts! The final result is the value 8, which is sent to the continuation $\alpha$. The reader may check that the machine given above
indeed yields this result when it is run on  the source code. (In later work, we intend to prove formally that this source machine is correctly simulated by the GAM.)

\smallskip
We give two further examples, in less detail.
The compilation of 
$$\letcbv{x}{y3}{\letcbv{v}{x4}{[\gamma]v}}
\quad[y\leftarrow(\lbd(z,\alpha).[\alpha](\lbd(u,\beta).[\beta](u+z)]$$
(assuming a binary adddition operation) is:

\smallskip
$\ppm{(?3,y)}{\_}\begin{branch} \pom{!\!\bullet} \ppm{?4}{0}\begin{branch}
\vdots\\
\pom{!m}\ppm{(!m,\gamma)}{\_}\\
\vdots
\end{branch}
\end{branch}$

\smallskip
$\begin{branch}
\vdots\\
\pom{(?n,y)}\ppm{!\bullet}{0}\begin{branch}
\vdots\\
\pom{?p}\:\ppm{!(n+p)}{0}\\
\vdots
\end{branch}\\
\vdots
\end{branch}$

\smallskip\noindent
with the following execution:

\smallskip
$\ppm{(?3,y)}{\_}\begin{branch} \mpom{!\bullet}{3} \ppm{?4}{0}\begin{branch}
\mpom{!7}{5}\ppm{(!7,\gamma)}{\_}
\end{branch}
\end{branch}$

\smallskip
$\begin{branch}
\mpom{(?3,y)}{2}\ppm{!\bullet}{0}\begin{branch}
\mpom{?4}{4}\:\ppm{!7}{0}
\end{branch}
\end{branch}$

\medskip\noindent
The compilation of 
$$\letcbv{x}{y(\lbd(z,\beta).[\beta]z)}{[\alpha]x}\quad
[y\leftarrow(\lbd(f,\gamma).\letcbv{u}{f3}{[\gamma]u})]$$
is:

\smallskip
$\ppm{(?\bullet,y)}{\_}\begin{branch}
\vdots\\
\pom{?n}\ppm{!n}{0}\\
\vdots\\
\pom{!p}\ppm{(!p,\alpha)}{\_}\\
\vdots
\end{branch}$

\smallskip
$\begin{branch}
\pom{(?\bullet,y)}\ppm{?3}{0}\begin{branch}
\vdots\\
\pom{!n}\ppm{!n}{1}\\
\vdots
\end{branch}
\end{branch}$

\smallskip\noindent
with the following execution:

\smallskip
$\ppm{(?\bullet,y)}{\_}\begin{branch}
\mpom{?3}{3}\ppm{!3}{0}\\
\mpom{!3}{5}\ppm{(!3,\alpha}{\_}
\end{branch}$

\smallskip
$\begin{branch}
\mpom{(?\bullet,y)}{2}\ppm{?3}{0}\begin{branch}
\mpom{!3}{4}\ppm{!3}{1}
\end{branch}
\end{branch}$

\medskip
We end the section by a short incursion into types. In figure \ref{CBV-types}, we present typing rules for
classical \pcf\ call-by-value trees, in a simple type system built over the base type $\iota$ of naturals, and over the function type:
$$\sigma :: \iota \mid (\sigma\rightarrow\sigma)$$
There is also a special type $\bot$, which is not used in compound types and serves only to type commands.  A typing context consists of a pair $\Gamma;\Delta$ of two multisets containing declarations $x:\sigma$ (ordinary variables) and $\alpha:\sigma$ (continuation variables), respectively.
We can type the successive states as closed judgements
$\vdash \coupe{c}{\rho}:\bot$,
with
$\Gamma;\Delta\vdash c:\bot$ for some $\Gamma$ and $\Delta$ such that for all $x:\sigma$ (resp. 
$\alpha:\tau$) appearing in $\Gamma$ (resp. $\Delta$), we have
$\vdash \rho(x):\sigma$ (resp. $\vdash\rho(\alpha):\tau$). The construct $\tilde\mu$ is typed as follows:
$$\seq{\Gamma,x:\tau;\Delta\vdash c:\bot}{\Gamma;\Delta\vdash \tilde\mu x.c:\tau}$$
Moreover, well-typed states are guaranteed to evolve without encountering the ``bad" termination case of section \ref{termination}. We omit the details, since we do not want to put emphasis on types here.

\begin{figure}
\Figbar
$$\begin{array}{c}
\seq{}{\Gamma;\Delta\vdash \ul{n}:\iota} \quad\quad 
\seq{}{\Gamma,x:\iota;\Delta\vdash x:\iota}\quad\quad
\seq{\Gamma,z:\sigma;\Delta,\beta:\tau\vdash c:\bot}
{\Gamma:\Delta\vdash \lbd(z,\beta).c:\sigma\rightarrow\tau}\\\\
\seq{\Gamma;\Delta,\alpha:\sigma\vdash V:\sigma}{\Gamma;\Delta,\alpha:\sigma\vdash[\alpha]V:\bot}\quad\quad\quad 
\seq{\qqs{i}{\Gamma,x:\iota;\Delta\vdash c_i:\bot}}
{\Gamma,x:\iota;\Delta\vdash \case x \:[\vec{\ul{n}}\rightarrow\vec{c}]:\bot}\\\\
\seq{\Gamma,y:\sigma\rightarrow \tau, x:\tau;\Delta\vdash c:\bot
\quad\quad\Gamma,y:\sigma\rightarrow \tau;\Delta\vdash V:\sigma}
{\Gamma,y:\sigma\rightarrow \tau;\Delta\vdash\letcbv{x}{yV}{c}:\bot}
\end{array}$$
\caption{Typed classical \pcf\ call-by-value trees}
\label{CBV-types}
\Figbar
\end{figure}

\subsection{Ludics} \label{Examples-Ludics}
Ludics \cite{GirLS} is a recent theory whose aim is to reconstruct logic on interactive principles.
While denotational semantics has been category-theory-oriented and type-oriented in the last 25 years, 
ludics deliberately adopts the view that computation is untyped, and that types may be built out of untyped objects as collections of such objects ``behaving the same way". The objects of ludics, called {\em designs}, are
derived from the skeletons of proofs in linear logic obtained after removing the type information (just retaining the subformula relations). 

\smallskip
In \cite{CuLL-ludintro}, we have shown that the designs can be presented by a syntax which is accessible without prior knowledge of linear logic and of the genesis of ludics:
$$\begin{array}{l}
M::=  \setc{J=\lbd\setc{x_j}{j\in J}.P_J}{J\in{\cal P}_f(\omega)}\\
P::=(x\cdot I)\setc{M_i}{i\in I}\mid \Omega \mid \demon
\end{array}$$

\noindent
Girard's original ludics enforces {\em affinity} assumptions (see \cite{CuLL-ludintro} for details).
Without these assumptions, this syntax reflects Maurel's exponential ludics  \cite{MaurelTh}.

\noindent The execution is driven by the following abstract machine.

$$\coupe{(x\cdot I)\setc{M_i}{i\in I}}{\rho}
\quad\longrightarrow\quad \coupe{P_I}{\rho_I\setc{x_i\leftarrow M_i[\rho]}{i\in I}}$$

\smallskip\noindent 
where $\rho(x)=\set{\ldots,I=(\lbd\setc{x_i}{i\in I}.P_I)[\rho_I],\ldots}$.
The reduction combines substitution of actual parameters (the $M_i$'s) for formal ones
(the $x_i$'s), typical of $\beta$-reduction, with a prior selection of a ``field" $I$.
In this framework, abstractions are replaced by a collection of abstractions, one for each
possible indexing set $I$. 
Whence our choice of a notation reminiscent of that used for records in
some object-oriented programming languages. 

We say that the evaluation {\em converges} when it reaches a stage $\coupe{\demon}{\rho}$, and that it diverges if either it reaches a stage $\coupe{\Omega}{\rho}$, or the computation never ends
(this cannot happen if both the design and the counter-design have finite depth, cf. section
\ref{termination}).

\medskip
It is easy to make this syntax fit into the format of abstract B\"ohm trees.
One first groups the two layers of indexing (by $i\in I$ and by $J\in{\cal P}_f(\omega)$). This gives
(by a textual transformation):
$$(x\cdot I)\setc{((i,J),(\lbd\setc{x_j}{j\in J}.P_{i,J}))}{i\in I,J\in{\cal P}_f(\omega)}$$
One then replaces $\setc{x_j}{j\in J}$ by ${\tt x}$ and every (bound) $(x_j.K)$ by $((j,K),{\tt x})$.
The two constants can be treated as special (terminal) free player's moves.

\section{The View Abstract Machine} \label{VAM-section}

In this section, we give a ``lighter" version of the GAM, where the state is a sequence of {\em moves}, called a {\em play} rather than a sequence of positions. The price to pay is that relevant information must be reconstructed
at each step on the fly. This version of the GAM is called the View Abstract Machine (VAM).

Before we define it formally, let us examine the play underlying the execution of the example
in section \ref{Examples-Boehm}. The following moves are played successively at steps $(1)$,
$(\appst{2})$ and $(2)$, $(\appst{3})$ and (3), etc...:
$\pom{\bullet}$, $\ppm{u}{\_}$ and $\pom{u}$, $\ppm{1}{0}$ and $\pom{1}$, etc... .
But the resulting sequence is not a position (but rather an interleaving of positions) of $\phi$ or $\psi$. At each step $(2n-1)$ (resp. $(2n)$), we have to reconstruct the appropriate position of $\phi$ (resp. $\psi$) -- the {\em view} at this step --, in order to apply $\phi$ (resp. $\psi$), as before. For example, the view at step $(2)$ consists of $\pom{u}$ only.

The syntax of VAM states is as follows:
$$\begin{array}{lll}
\mbox{VAM states:} &&\Gamma ::= \litem{\bullet}{1} \mid  \Gamma\litem{\ppm{a}{\kappa}}{\appst{n}} \mid \Gamma\litem{ a}{2n} \mid   \Gamma\litem{\jitem{a}{\appst{m}}}{n}
\end{array}$$

\noindent
We next define  {\em jumps} and {\em views}, which are our tools for reconstructing information:

$$\seq{\position{\Gamma}{n}=\jitem{a}{\appst{m}}}{\jump{\Gamma}{}{n}=m-1}$$

$$\seq{}
{\view{\Gamma}{1}=\bullet}\quad\quad\seq{\position{\Gamma}{(2n)}=a}
{\view{\Gamma}{2n}=a}\quad\quad\seq{\position{\Gamma}{n}=\jitem{a}{\appst{m}}}
{\view{\Gamma}{n}=\view{\Gamma}{m-
1}\:(\position{\Gamma}{\appst{m}})\:a}$$

\noindent 
The rules of the VAM are given in figure \ref{VAM-rules}.

\begin{figure}
\Figbar
$$\seql{$(1)$}{}{\mapsto\litem{\bullet}{1}}$$

$$\seql{$(\appst{2n})$}{\phi(\view{\Gamma}{2n-1})=\ppm{a}{\kappa}}
{\Gamma\mapsto\Gamma\litem{\ppm{a}{\kappa}}{\appst{2n}}}$$

$$\seql{$(2n)_b$}{\hd{\Gamma}=\litem{\ppm{a}{i}}{\appst{2n}}\quad
\jump{\Gamma}{i}{2n-1}=2m-1}
{\Gamma\mapsto\Gamma\litem{\vitem{a}{\appst{2m-1}}}{2n}}$$

$$\seql{$(2n)_f$}{\hd{\Gamma}=\litem{p\ppm{a}{\_}}{\appst{2n}}}
{\Gamma\mapsto\Gamma\litem{a}{2n}}$$

$$\seql{$(\appst{2n+1})$}{\psi(\view{\Gamma}{2n})=\ppm{a}{\kappa}}
{\Gamma\mapsto\Gamma\litem{\ppm{a}{\kappa}}{\appst{2n+1}}}$$

$$\seql{$(2n+1)$}{\hd{\Gamma}=\litem{\ppm{a}{i}}{\appst{2n+1}}\quad
\jump{\Gamma}{i}{2n}=2m}
{\Gamma\mapsto\Gamma\litem{\vitem{a}{\appst{2m}}}{2n+1}}$$
\caption{The View Abstract Machine (VAM)}
\label{VAM-rules}
\Figbar
\end{figure}

\medskip
We collect here a few observations.
\begin{itemize}
\item Most works on game semantics do not define strategies as sets of views like we do.
For example, Hyland and Ong's interpretation of $(u(\lbd x. u(\lbd y.x))$ (cf. section \ref{Examples-Boehm}) contains the position 
$$\bullet\:\ppm{u}{\_}\:
\ul{1\:\ppm{u}{\_}\:1\:\ppm{1}{1}}
\:1\ppm{u}{\_}$$
whose view  $\bullet\:\ppm{u}{\_}\:
1\:
\ppm{u}{\_}$ is obtained by removing the underlined portion.  There are some advantages to this ``plethorous" definition of strategies:
\begin{itemize}
\item  In such positions, there is no dissymetry between Player and Opponent, since one has broken
the requirement that Opponent always plays just below Player (this is why there are no pointers originating from opponent's moves in abstract B\"ohm trees). 
\item The composition of strategies can be defined algebraically in one line as ``composition + hiding"
\cite{HO,AJM}.
\end{itemize}
The drawback is that it gives an infinite representation of finite objects, since, say, also
$$\bullet\:\ppm{u}{\_}\:
1\:
\ppm{u}{\_}\:
1\:\ppm{1}{1}\:1\ppm{u}{\_}\:1\ppm{u}{\_}$$
etc... belong to the strategy.  The existence of these two kinds of representation of a strategy -- the sober one, isomorphic to the underlying term, and the plethoric one -- was first recognized by Felscher, who calls them $E$-dialogue and $D$-dialogue, respectively \cite{Felscher}.
(Felscher's work continues a school of thought initiated by Lorenz and Lorenzen, where proofs are viewed as a dialogue between a defendant and an opponent, a bit like in a PhD defence. Unfortunately, these works emphasized provability rather than proofs.)

\item The translation from a VAM state to a GAM state (see section \ref{VAM-GAM}) can be understood as a desenquentialization process: the play $\Gamma$ gets translated to a collection of views (with repetitions) -- a thick tree of views.
The equivalence between the GAM and the VAM (again, see section \ref{VAM-GAM})  guarantees that the information collected in the multiplexed opponent's moves of the translation contains all the original sequential information, implicitly.
 However, by moving to the tree isomorphism class of $\trad{\Gamma}{VG}$, this information is really lost, and then it makes sense to talk about desequentialization. This forgetful desequentialization is close in spirit to the work of Boudes \cite{Boudes2005} on the analysis of the relation between game semantics and coherence semantics (a more traditional kind of model).

\item
 The concept of multiplexing is
not only relevant for dynamic issues, but also for static ones. In a typed setting,
each position in a strategy encodes a multiplexed traversal of its
type.  
The two kinds of multiplexing obey a dual discipline: the dynamic one 
multiplexes the
opponent's moves, whereas the static one multiplexes the player's moves.
We illustrate this with a simple example:
the most general type of 
$M=(\lbd u.u(\lbd x. u(\lbd y.x)))$, in Hindley's sense, is
$$((X\fun Y)\fun Y)\fun Y$$
which we represent as
$$Y\begin{branch} Y\begin{branch} Y\begin{branch} X\end{branch}\end{branch}
\end{branch}$$
We show how $M$, which consists of one path only,
encodes a multiplexing of its type:

$$\comm{(\lbd u}{Y}\begin{branch} \comm{u_1}{\mpom{Y}{2}}
\begin{branch} \comm{(\lbd x}{Y}\begin{branch}\comm{x}{\mpom{X}{6}}\end{branch}
\end{branch}\\
\comm{u_2}{\mpom{Y}{4}}\begin{branch}\comm{(\lbd y}{}\end{branch}
\end{branch}$$

We shall make a timid incursion in linear logic, by suggesting that this
multiplexing of types can be understood
as rewriting of trees by means of the rule $!A\fun(!A\otimes A)$.
For example, the type $((X\fun Y_3)\fun Y_2)\fun Y_1$, expressed 
in linear logic ($\fun$ being
now the linear arrow), is:
$$!(!(!X\fun Y_3)\fun Y_2)\fun Y_1$$
and can be rewritten to:
$$(!(!(!X\fun Y_3)\fun Y_2)\otimes !(!X\fun Y_3)\fun Y_2)\fun Y_1$$
which corresponds to the multiplexing above.
\end{itemize}

\section{A concrete version of the GAM} \label{SAM-section}

In this section, we complete the operational picture by introducing two other formulations of the GAM:
the Strategic Abstract Machine (SAM), and its concrete version, the Environment Abstract Machine (EAM), formulated in terms of the concrete syntax of section \ref{syntax-abt}.

\subsection{Strategic abstract machine} \label{SAM-subsection}

Instead of marking the multiplexed nodes with numbers $n$, which in their 
form $\appst{n}$ point us to player's positions in the other strategy, we could 
directly 
mark them with these positions. 
This yields the following version of the GAM:

$$\seq{}{\mapsto\litem{\bullet}{1}}$$

$$\seq{\hd{\Gamma}=\litem{p}{2n-1}\quad
\phi(\erase{p})=\ppm{a}{i}}
{\Gamma\mapsto\Gamma\litem{p\ppm{a}{i}}{\appst{2n}}}$$

$$\seq{\hd{\Gamma}=\litem{p\ppm{a}{i}}{\appst{2n}}\quad
\dnumber{\pop^i(p)}=q}
{\Gamma\mapsto\Gamma\litem{q\vitem{a}{p\ppm{a}{i}}}{2n}}$$

$$\seq{\hd{\Gamma}=\litem{p\ppm{a}{\_}}{\appst{2n}}}
{\Gamma\mapsto\Gamma\litem{\vitem{a}{p\ppm{a}{\_}}}{2n}}$$

$$\seq{\hd{\Gamma}=\litem{p}{2n}\quad
\psi(\erase{p})=\ppm{a}{i}}
{\Gamma\mapsto\Gamma\litem{p\ppm{a}{i}}{\appst{2n+1}}}$$

$$\seq{\hd{\Gamma}=\litem{p\ppm{a}{i}}{\appst{2n+1}}\quad
\dnumber{\pop^i(p)}=q}
{\Gamma\mapsto\Gamma\litem{q\vitem{a}{p\ppm{a}{i}}}{2n+1}}$$

\medskip
\noindent
The distinction between $n$ and $\appst{n}$  becomes then useless, but we keep it in order to stay with the same number of rules.

\medskip


%

But now we observe that the machine, 
expressed in this format, is ``history-free''.
We can thus remove $\Gamma$ altogether, and describe the machine as a
rewriting system on (nested)  positions:
$$\begin{array}{llccc}
&& n && \appst{n}\\\\
\mbox{SAM states} &&  {\tt q}::=\bullet\mid\vitem{a}{{\tt r}}\mid{\tt r'}\vitem{a}{{\tt r}} &&
{\tt r}::={\tt q}\ppm{a}{\kappa}
\end{array}$$

We arrive at the machine described in figure \ref{SAM-rules}, called here Strategic Abstract Machine
(SAM).

\begin{figure}
\Figbar
$$\seql{$(1)$}{}{\hfs{1}{\bullet}}$$

$$\seql{$(\appst{2n})$}{\phi(\erase{{\tt q}})=\ppm{a}{\kappa}}
{{\tt q}\hfs{\appst{2n}}{\tt q}\ppm{a}{\kappa}}$$

$$\seql{$(2n)_b$}{\dnumber{\pop^i({\tt q})}={\tt r'}}
{{\tt q}\ppm{a}{i}\hfs{2n}{\tt r'}\vitem{a}{{\tt q}\ppm{a}{i}}}$$

$$\seql{$(2n)_f$}{}
{{\tt q}\ppm{a}{\_}\hfs{2n}\vitem{a}{{\tt q}\ppm{a}{\_}}}$$

$$\seql{$(\appst{2n+1})$}{\psi(\erase{{\tt q}})=\ppm{a}{\kappa}}
{{\tt q}\hfs{\appst{2n+1}}{\tt q}\ppm{a}{\kappa}}$$

$$\seql{$(2n+1)$}{\dnumber{\pop^i({\tt q})}={\tt r'}}
{{\tt q}\ppm{a}{i}\hfs{2n+1}{\tt r'}\vitem{a}{{\tt q}\ppm{a}{i}}}$$
\caption{The Strategic Abstract Machine (SAM)}
\label{SAM-rules}
\Figbar
\end{figure}

\subsection{The Environment Abstract Machine} \label{EAM-machine}

The SAM can be formulated in terms of the concrete syntax of section \ref{syntax-abt}, and in this form, it appears as the natural generalization of (a stack-free version of) Krivine abstract machine. We call it the Environment Abstract Machine. 
The states of the EAM are pairs (code, environment), recursively defined as follows:

$$\begin{array}{lllllll}
\mbox{Closures:} && M[\rho]\\
\mbox{Environments:} && \mbox{maps $\rho$ from variables to closures}\\
\mbox{EAM states:} && \coupe{P}{\rho}
\end{array}$$

\noindent
The rules of the EAM are given in figure \ref{EAM-rules}. 

\begin{figure}
\Figbar
$$\seq{\rho(\cppm{a}{\star})=(\lbd{\tt y}.Q)[\rho']}
{\coupe{\cppm{a}{\star}\:\setc{(b,M_b)}{b\in B}}{\rho}  \quad\longrightarrow\quad  \coupe{Q} {\rho'[\cppm{b}{{\tt y}}\leftarrow M_b[\rho]\:|\:b\in B]}}$$

$$\seq{\rho(\cppm{a}{\star})=(Q)[\rho']}
{\coupe{\cppm{a}{\star}\:\set{\:}}{\rho}  \quad\longrightarrow\quad  \coupe{Q}{\rho'}}$$

\caption{The Environment Abstract Machine (EAM)}
\label{EAM-rules}
\Figbar
\end{figure}

\section{Strong reduction} \label{Strong-section}

In this section, 
we show how to extend the
GAM to a {\em strong} machine, which computes a 
complete strategy. The strong machine which we present can build on demand, in a {\em stream-like} fashion, any position of
the composition of $\phi$ and $\psi$. The initial step of the machine
builds the initial position $\bullet$ of the composition.
Either of the (two first) terminating  states discussed in section \ref{termination}, when first encountered,
produces a player's move in the composition that answers the initial
query $\bullet$: in the first case this player's move is written
$\sppm{a}{1}$, in the second case, it is written as usual $\ppm{a}{\_}$.
Then the pilot of the strong machine may decide to raise a new query,
expressed as a position
$\bullet\sppm{a}{1}\mpom{b}{2n}$ (in the first case)
or $\bullet\ppm{a}{\_}\mpom{b}{2n+1}$ (in the second case). Note that
the alternation is broken: this new step, which is of the form $(2n)$ in the first case,
and $(2n+1)$ in the second case, is performed in the same strategy as the
previous step. 

\smallskip
The machine states consist now of triplets of the form:
\begin{itemize}
\item[-]
$\state{\Gamma}{?}{q}$, where $\Gamma$ is as before, and where $?$
witnesses that we are working on answering a query $q$,
\item[-]
$\state{\Gamma}{!}{r}$, where $\Gamma$ is as before,  and where $!$ witnesses
that the query at the previous stage is just answered.
\end{itemize}
The complete set of rules of the strong GAM is given in figure \ref{Strong-GAM-rules}.
The rules $(1)^s$,
$(\appst{n})_{\phi}^s$,
$(n)_{b,\phi}^s$, $(n)_{f,\phi}^s$, $(\appst{n})_{\psi}^s$, and
$(n)_{\psi}^s$, are just rephrasings of the rules of the (weak) GAM.

\begin{figure}
\Figbar
$$\begin{array}{c}
\seql{$(1)^s$}{}{\mapsto\state{\slitem{\mpom{\bullet}{1}}{1}{\phi}}{?}
{\spom{\bullet}{1}}}\quad\quad \seql{$(n)_{f,\phi}^s$}{\hd{\Gamma}=\slitem{\dyn{q}\ppm{a}{\_}}{\appst{n}}{\phi}}
{\state{\Gamma}{?}{r}\mapsto\state{\Gamma\slitem{\mpom{a}{n}}{n}{\psi}}{?}{r}}\\\\
\seql{$(\appst{n})_{\phi}^s$}{\hd{\Gamma}=\slitem{\dyn{q}}{n-1}{\phi}\quad
\phi(\erase{\dyn{q}})=\ppm{a}{i}}
{\state{\Gamma}{?}{r}\mapsto
\state{\Gamma\slitem{\dyn{q}\ppm{a}{i}}{\appst{n}}{\phi}}{?}{r}}\\\\
\seql{$(n)_{b,\phi}^s$}{\hd{\Gamma}=\slitem{\dyn{q}\ppm{a}{i}}{\appst{n}}{\phi}\quad
\dnumber{\pop^i(\dyn{q})}=m\quad
\position{\Gamma}{\appst{m}}=\dyn{r'}\in\psi}
{\state{\Gamma}{?}{r}\mapsto
\state{\Gamma\slitem{\dyn{r'}\mpom{a}{n}}{n}{\psi}}{?}{r}}\\\\
\seql{$(\appst{n})_{\psi}^s$}{\hd{\Gamma}=\slitem{\dyn{q}}{n-1}{\psi}\quad
\psi(\erase{\dyn{q}})=\ppm{a}{i}}
{\state{\Gamma}{?}{r}\mapsto
\state{\Gamma\slitem{\dyn{q}\ppm{a}{i}}{\appst{n}}{\psi}}{?}{r}}\\\\
\seql{$(n)_{\psi}^s$}{\hd{\Gamma}=\slitem{\dyn{q}\ppm{a}{i}}{\appst{n}}{\psi}\quad
\dnumber{\pop^i(\dyn{q})}=m\quad
\position{\Gamma}{\appst{m}}=\dyn{r'}\in\phi}
{\state{\Gamma}{?}{r}\mapsto\state{\Gamma\slitem{\dyn{r'}\mpom{a}{n}}{n}{\phi}}{?}{r}}\\\\
\seql{$(!)_{\phi}^s$}{\hd{\Gamma}=\slitem{\dyn{q}\ppm{a}{i}}{\appst{n}}{\phi}\quad
\dnumber{\pop^i(\dyn{q})}=m\quad
(m=1\mor\position{\Gamma}{\appst{m}}\in\phi)}
{\state{\Gamma}{?}{r}\mapsto\state{\Gamma}{!}{r\sppm{a}{m}}}\\\\
\seql{$(!)_{b,\psi}^s$}{\hd{\Gamma}=\slitem{\dyn{q}\ppm{a}{i}}{\appst{n}}{\psi}\;\;
\dnumber{\pop^i(\dyn{q})}=m\;\;
\position{\Gamma}{\appst{m}}\in\psi}
{\state{\Gamma}{?}{r}\mapsto\state{\Gamma}{!}{r\sppm{a}{m}}}\quad
\seql{$(!)_{f,\psi}^s$}{\hd{\Gamma}=\slitem{\dyn{q}\ppm{a}{\_}}{\appst{n}}{\psi}}
{\state{\Gamma}{?}{r}\mapsto\state{\Gamma}{!}{r\sppm{a}{\_}}}\\\\
\seql{$(?)_{\chi}^s$}{\hd{\Gamma}=\slitem{\dyn{r}'}{\appst{n}}{\phi}\;\;
\erase{\dyn{r}'}\pom{a}\in\dom{\chi}\;\; (\chi=\phi\mbox{ or }\psi)}
{\state{\Gamma}{!}{r}\mapsto
\state{\Gamma\slitem{p\mpom{a}{n}}{n}{\chi}}{?}{\dyn{r}'\spom{a}{n}}}
\end{array}$$
\caption{The Strong GAM}
\Figbar
\label{Strong-GAM-rules}
\end{figure}

\smallskip
We have been a bit vague about the syntax of the positions appearing 
as the third component of the states of the machine. The point is that
the dynamics of the computation does not allow us to immediately extract 
pointers encoded as offsets like those we had for the positions of $\phi$,
$\psi$ we started with. Instead, we have pointers in the form of addresses:
when a position $q\sppm{a}{m}$ is reached (at a step $(!)_{\phi}^s$
or $(!)_{b,\psi}^s$), we mean that the move $\sppm{a}{m}$ points to the unique
opponent's move of $q$ of the form $\spom{a}{m}$.

We have already remarked that when going from ``weak''  to ``strong'', the nice
alternation between $\phi$ and $\psi$ is broken (rules $(?)_{\phi}^s$ and
$(?)_{\psi}^s$). This  could be one abstract definition
of weak evaluation: evaluation is weak as long as it strictly alternates 
between the strategy and the counter-strategy.

The composition of $\phi$ and $\psi$ is defined as the collection of the positions obtained through the various non-deterministic executions of the GAM (recompiled in order to obtain standard abstract B\"ohm tree positions). We obtain in this way a category (actually, a multicategory, where morphisms have a sequence or set of objects as domain and an object as codomain) of strategies (see \cite{Cur97}).
The identities, which in the multicategorical setting are in fact projections (from the domain to one of the objects of the domain), are so-called copy-cat or  fax strategies (see section \ref{eta-section}). The associativity of the composition amounts to define a three-way machine that lets all the strategies involved interact.

\medskip
We give two examples which illustrate the use of the new rules:

\medskip
\noindent
Example 1
\begin{itemize}
\item[] 
Function: the strategy for $(\lbd u.x(\lbd v.u(v)))$ is:

$$\comm{(\lbd u}{\bullet}\ppm{x}{\_}\begin{branch}
\comm{(\lbd v}{\pom{1}}\comm{u}{\ppm{1}{1}}\begin{branch}
\comm{(}{\pom{1}}\comm{v}{\ppm{1}{1}}
\end{branch}
\end{branch}$$

\item[]
Argument: the strategy for $\Sub{x}{(\lbd r.r(z))}$ is:

$$\comm{(\lbd r}{\pom{x}}\comm{r}{\ppm{1}{0}}\begin{branch}
\comm{(}{\pom{1}}\ppm{z}{\_}
\end{branch}$$

\item[]
Function multiplexed tree:

$$\comm{(\lbd u\;?}{\mpom{\bullet}{1}}\ppm{x}{\_}\begin{branch}
\mpom{1}{3}\comm{u\;!_{\phi}}{\ppm{1}{1}}\begin{branch}
\comm{(\;?_{\phi}}{\mpom{1}{4}}\ppm{1}{1}
\end{branch}
\end{branch}$$

\item[]
Argument multiplexed tree:

$$\mpom{x}{2}\ppm{1}{0}\begin{branch}
\mpom{1}{5}\comm{!_{f,\psi}}{\ppm{z}{\_}}
\end{branch}$$
\item[] Reading back of the composition: $(\lbd u.u(z))$.
\end{itemize}

\medskip
\noindent
Example 2
\begin{itemize}
\item[]
Function: the strategy for $(x(\lbd u.u))$ is:

$$\comm{(}{\bullet}\ppm{x}{\_}\begin{branch}
\comm{(\lbd u}{\pom{1}}\comm{u}{\ppm{1}{0}}
\end{branch}$$

\item[]
Argument: the strategy for $\Sub{x}{(\lbd z.y(\lbd t.z(t)))}$ is:

$$\comm{(\lbd z}{x}\ppm{y}{\_}\begin{branch}
\comm{(\lbd t}{\pom{1}}\comm{z}{\ppm{1}{1}}\begin{branch}
\comm{(}{\pom{1}}\comm{t}{\ppm{1}{1}}
\end{branch}
\end{branch}$$

\item[]
Function multiplexed tree:

$$\comm{(\;?}{\mpom{\bullet}{1}}\ppm{x}{\_}\begin{branch}
\mpom{1}{4}\ppm{1}{0}
\end{branch}$$

\item[]
Argument multiplexed tree:

$$\mpom{x}{2}\comm{!_{f,\psi}}{\ppm{y}{\_}}\begin{branch}
\comm{(\lbd t\;?_{\psi}}{\mpom{1}{3}}\ppm{1}{1}\begin{branch}
\mpom{1}{5}\comm{t\;!_{b,\psi}}{\ppm{1}{1}}
\end{branch}
\end{branch}$$
\item[-] 
Reading back of the composition: $(y(\lbd t.t))$.
\end{itemize}

\section{Evaluating non-normal forms} \label{non-normal}

The syntactic formulation of abstract B\"ohm trees (cf. section \ref{syntax-abt}) suggests us to extend the syntax to ``non-normal forms", as follows:
$$\begin{array}{l}
P::= M\:\setc{(b,M_b)}{b\in B}\;\;\mid\;\; \cppm{a}{\star}\:\setc{(b,M_b)}{b\in B}  \quad(\star={\tt x}\mbox{ or } \star=\_)\\
M::=(\lbd {\tt x}.P)
\end{array}$$
and to equip the language with the following notion of reduction (a generalized version of the $\beta$-rule of the $\lbd$-calculus):
$$\begin{array}{l}
(\lbd {\tt x}.P) \:\setc{(b,M_b)}{b\in B} \quad\longrightarrow \quad P\:[\ppm{b}{{\tt x}}\leftarrow M_b\:|\:b\in B]
$$
\end{array}$$

\noindent
The machinery of the GAM can be extended to such terms.
Expressed in terms of $\lbd$-calculus, 
the basic idea, which is somewhat folklore, is to replace any redex
$(\lbd x.M)N$ by an indirection of the form $(\star(N))\Sub{\star}{(\lbd x.M)}$,
where $\star$ is a special variable name.  
We make the following ajustments:
\begin{enumerate}
\item We compile the terms in two steps. First we obtain a term of the core syntax of normal forms, where we have reserved a second special move $\star$:
$$\trad{M\:\set{\ldots,(b,M_b),\ldots}}{\star}=
(\ppm{\star}{\_}\:\set{(\star,M),\ldots,(b,M_b),\ldots}$$
And then we compile the core term.
\item We add the following rule to the GAM:

$$\seql{$(n)_{\star}$}{\hd{\Gamma}=\litem{\dyn{q}\ppm{\star}{\_}}{\appst{n}}}
{\Gamma\mapsto\Gamma\litem{\dyn{q}\ppm{\star}{\_}\mpom{\star}{n}}{n}}$$

\noindent
In addition, the rules $(\appst{2n})$ and $(\appst{2n+1})$ collapse in a single rule $(\appst{n})$, since there is now a single strategy $\chi$  interacting internally with itself. Similarly, the rules $(2n)_b$ and $(2n+1)$ become just one rule $(n)$, and the rule $(2n)_f$ disappears. 
The changes are summarized in figure \ref{GAM*-rules}, and the resulting machine is called the $\mbox{GAM}^\star$.

\begin{figure}
\Figbar

$$\seql{$(1)$}{}{\mapsto\litem{\mpom{\bullet}{1}}{1}}$$

$$\seql{$(\appst{n})$}{\hd{\Gamma}=\litem{\dyn{q}}{n-1}\quad
\chi(\erase{\dyn{q}})=\ppm{a}{i}}
{\Gamma\mapsto\Gamma\litem{\dyn{q}\ppm{a}{i}}{\appst{n}}}$$

$$\seql{$(n)$}{\hd{\Gamma}=\litem{\dyn{q}\ppm{a}{i}}{\appst{n}}\quad
\dnumber{\pop^i(\dyn{q})}=m\quad
 \position{\Gamma}{\appst{m}}=\dyn{r}'}
{\Gamma\mapsto\Gamma\litem{\dyn{r}'\mpom{a}{n}}{n}}$$

$$\seql{$(n)_{\star}$}{\hd{\Gamma}=\litem{\dyn{q}\ppm{\star}{\_}}{\appst{n}}}
{\Gamma\mapsto\Gamma\litem{\dyn{q}\ppm{\star}{\_}\mpom{\star}{n}}{n}}$$
\caption{The $\mbox{GAM}_\star$}
\label{GAM*-rules}
\Figbar
\end{figure}

\end{enumerate}
With this machine we can accommodate $\lbd$-terms. We illustrate the $\mbox{GAM}^\star$ 
 with a simple example, which is a variation on the theme
of static binding. Let
$$M=((\lbd x.(\lbd f.(\lbd x.f(t))(\lbd z.u))(\lbd y.x(y)))(\lbd z.z))$$
which is easier to read with syntactic sugar:
$$M=(\mbox{let }x=(\lbd z.z)\mbox{ in }(\mbox{let }f=(\lbd y.x(y))\mbox{ in }
(\mbox{let }x=(\lbd z.u)\mbox{ in }(f(t))
)))$$
It illustrates static binding: the final result is $t$, not $u$, i.e., the
relevant value for $x$ is $(\lbd z.z)$, which was the value of $x$ at the time
of declaration of $f$,
and not $(\lbd z.u)$ which is the execution time value of $x$ when $f$ is
called.

We can define the compilation of $\lbd$-terms by factoring through the term notation of section \ref{syntax-abt}. One adds the following obvious clause in the translation of section \ref{Examples-Boehm}:
$$\trad{MB_1\ldots B_n}{}=M\set{(1,\trad{B_1}{}),\ldots,(n,\trad{B_n}{})}$$
Here is the full compilation of $M$:
$$\comm{(}{\bullet}\ppm{\star}{\_}\begin{branch}
\comm{(\lbd z}{\pom{1}}\comm{z}{\ppm{1}{0}}\\
\comm{(\lbd x}{\pom{\star}}\ppm{\star}{\_}\begin{branch}
\comm{(\lbd y}{\pom{1}}\comm{x}{\ppm{1}{1}}\begin{branch}
\comm{(}{\pom{1}}\comm{y}{\ppm{1}{1}}\end{branch}\\
\comm{(\lbd f}{\pom{\star}}\ppm{\star}{\_}\begin{branch}
\comm{(\lbd z}{\pom{1}}\ppm{u}{\_}\\
\comm{(\lbd x}{\pom{\star}}\comm{f}{\ppm{1}{1}}\begin{branch}
\comm{(}{\pom{1}}\ppm{t}{\-}
\end{branch}
\end{branch}
\end{branch}
\end{branch}$$  
We now execute this term, using the acquired skills. 

\medskip
Multiplexed (compilation of) $M$:
$$\comm{(}{\mpom{\bullet}{1}}\ppm{\star}{\_}\begin{branch}
\comm{(\lbd z}{\mpom{1}{6}}\comm{z}{\ppm{1}{0}}\\
\comm{(\lbd x}{\mpom{\star}{2}}\ppm{\star}{\_}\begin{branch}
\comm{(\lbd y}{\mpom{1}{5}}\comm{x}{\ppm{1}{1}}\begin{branch}
\comm{(}{\mpom{1}{7}}\comm{y}{\ppm{1}{1}}\end{branch}\\
\comm{(\lbd f}{\mpom{\star}{3}}\ppm{\star}{\_}\begin{branch}
\comm{(\lbd z}{\pom{1}}\ppm{u}{\_}\\
\comm{(\lbd x}{\mpom{\star}{4}}\comm{f}{\ppm{1}{1}}\begin{branch}
\comm{(}{\mpom{1}{8}}\ppm{t}{\-}
\end{branch}
\end{branch}
\end{branch}
\end{branch}$$

\section{Evaluating and separating untyped $\lbd$-terms} \label{eta-section}

Consider the following very simple example, expressed in concrete syntax:
$$\coupe{\cppm{x}{\_}\set{\:}}{[x\leftarrow (\lbd{\tt y}.\cppm{a}{{\tt y}}]}$$
whose compilation in terms of trees of positions is:

\medskip
$\begin{array}{l}
\bullet \ppm{x}{\_}\\\\
x\ppm{a}{0}
\end{array}$

\smallskip\noindent
The computation is blocked at step $(\appst{3})$ (third case of termination, cf. section \ref{termination}):

\medskip
$\begin{array}{l}
\mpom{\bullet}{1} \ppm{x}{\_}\\\\
\mpom{x}{2}\ppm{a}{0}
\end{array}$

\medskip
\noindent The problem is that $x$ should have an argument. Evaluation can proceed if we perform an ${\eta}$-expansion of $x$. Recall that in the $\lbd$-calculus, the $\eta$-rule asserts
$$\lbd x.Mx=x$$
 for all $M$ and $x$ such that $x$ does not occur freely in $M$. In the present setting, we need to replace $(\cppm{x}{\_})$ by $(\lbd{\tt z}.\cppm{x}{\_}\set{(a,(\cppm{1}{{\tt z}}))}$. This looks simple enough, but, in fact, we cannot really do this for arbitrary abstract B\"ohm trees.
 Consider
 $\lbd x_1\ldots x_p.yM_1\ldots M_q$
 which $\eta$-expands to 
 $$\lbd x_1\ldots x_pz.yM_1\ldots M_qz\;.$$
 In compiled form, we have that 
 $\lbd {\tt x}.\cppm{y}{\_}\set{(1,B_1),\ldots,(q,B_q)}$ expands to
$$\lbd {\tt x}.\cppm{y}{\_}\set{(1,B_1),\ldots,(q,B_q), (q+1,(\cppm{p+1}{{\tt x}}))}$$
The point is that we need to name (or give an address to) the two moves which have been added: the new opponent's move has address $q+1$ since $z$ is to be the $(q+1)$-th argument of 
$y$, and the new player's move has address $p+1$ since it is to be the $(p+1)$--th abstracted variable.
This supposes a notion of  sequencing among arguments and abstracted variables, which is present in the examples treated in sections \ref{Examples-Boehm},  \ref{Examples-Pcf}, \ref{Examples-Classical}, and \ref{Examples-CBV}, but was ``lost in translation". 
In particular, in order to evaluate untyped $\lambda$-terms, we need to record more information during the compilation (section \ref{GAM-eta}).  

The $\eta$-expansion plays an essential role in the proof of an important theorem of the $\lbd$-calculus due to Corrado B\"ohm. We briefly recall what this theorem is about, and illustrate it through an example 
(section \ref{xy-x(xy)}).

There is however one instance where $\eta$-expansion makes sense for arbitrary B\"ohm trees: when the sets of arguments and abstracted variables are empty. This special case is very important, since it provides us with the missing bit of section \ref{Strong-section}: the projections, or copy-cat (or {\em fax}, in the nice terminology of \cite{GirLS}) strategies (section \ref{fax}).   We conclude with a discussion  contrasting B\"ohm's theorem  with the separation theorem of ludics (section \ref{Discussion-separation}).

\subsection{Incorporating $\eta$ into the GAM} \label{GAM-eta}

We redefine the compilation of B\"ohm trees,
keeping now 
the number of abstracted variables (resp. the number of arguments)
as a superscript on opponent's moves (resp. on player's moves):

\medskip
\noindent Abstract syntax:
$$\begin{array}{l}
P::= \ppm{a}{\kappa}^m\:\set{(1^{n_1},M_1),\ldots,(m^{n_m},M_m)}\\ 
M::=(P)
\end{array}$$

\noindent Concrete syntax:
$$\begin{array}{l}
P::= \cppm{a}{\star}^m\:\set{(1,M_1),\ldots,(m,M_m)} \\
M::=(\lbd {\tt x}^n.P)
\end{array}$$

\noindent
Compiling from untyped $\lbd$-calculus to concrete syntax:
$$\begin{array}{l}
\trad{xB_1\ldots B_m}{}=\cppm{x}{\_}^m\set{(1,\trad{B_1}{},\ldots,(m,\trad{B_m})}\\
\trad{{\tt x}_iB_1\ldots B_m}{}=\cppm{i}{{\tt x}}^m\set{(1,\trad{B_1}{},\ldots,(m,\trad{B_m})}
\\\\
\trad{\lbd {\tt x}_1,\ldots, {\tt x}_n.P}{}=(\lbd{\tt x}^n.\trad{P}{})
\end{array}$$

\noindent
Compiling from concrete syntax to abstract syntax:
 $$\begin{array}{l}
 \trad{(\lbd {\tt x}^n.P)}{ca}_L=(\trad{P}{ca}_{{\tt x}\cons L})^n\\\\
 \trad{\cppm{a}{\_}^m\:\setc{(1,M_1),\ldots,(m,M_m)}}{ca}_L\\
 \quad=\ppm{a}{\_}^m\:\set{\coerce{(1,\trad{M_1}{ca}_L)},\ldots,\coerce{(m,\trad{M_m}{ca}_L)}}\\
  \trad{\cppm{a}{{\tt x}}^m\:\setc{(1,M_1),\ldots,(m,M_m)}}{ca}_L\\
  \quad=\ppm{a}{L_{{\tt x}}}\:\set{\coerce{(1,\trad{M_1}{ca}_L)},\ldots,\coerce{(m,\trad{M_m}{ca}_L)}}
 \end{array}$$

\noindent where $\coerce{(i,M^n}=(i^n,M)$.

\smallskip The execution is driven by a variant of the GAM, which  we call $\mbox{GAM}_\eta$.
The $\mbox{GAM}_\eta$ continues to manipulate positions which do not have superscripts.
But the function which maps the positions of $\phi$ to the same positions where the superscripts have been erased is injective. We use the notation ${\it filter}(\dyn{r},\phi)$ to denote the inverse to this injection (the execution keeps within the range of the injection).

\smallskip
The $\eta$-expansion process is taken care of by the following additional rules:
$$\seql{$(\ol{2n})_\eta$}{\begin{array}{l}\hd{\Gamma}=\litem{\dyn{q}}{2n-1}\quad\dyn{q}=\dyn{r}\mpom{m}{2n-1}\\{\it filter}(\dyn{r},\phi)=r_1a^{n_
1}\ppm{b}{\kappa}^{n_2}
\quad m>n_2
\end{array}}
{\begin{array}{l}\Gamma\mapsto\Gamma\litem{\dyn{q}\ppm{m-n_2+n_1}{1}}{\ol{2n}}\\
(\Phi,\Psi)\mapsto(\Phi\union\set{{\it filter}(\dyn{r},\phi)m^0\ppm{m-n_2+n_1}{1}^0},\Psi)
\end{array}}$$

$$\seql{$(\ol{2n+1})_\eta$}{\begin{array}{l}\hd{\Gamma}=\litem{\dyn{q}}{2n}\quad\dyn{q}=\dyn{r}\mpom{m}{2n-1}\\{\it filter}(\dyn{r},\psi)=r_1a^{n_
1}\ppm{b}{\kappa}^{n_2}
\quad m>n_2
\end{array}}
{\begin{array}{l}\Gamma\mapsto\Gamma\litem{\dyn{q}\ppm{m-n_2+n_1}{1}}{\ol{2n+1}}\\
(\Phi,\Psi)\mapsto(\Phi,\Psi\union\set{{\it filter}(\dyn{r},\phi)m^0\ppm{m-n_2+n_1}{1}^0})
\end{array}}$$

\noindent Notice that $\Phi$ and $\Psi$ are progressively updated: they are $\eta$-expanded, as the need arises.

\medskip
We can now complete the execution of the example of the preamble of this section:

\medskip
$\begin{array}{l}
\mpom{\bullet}{1} \ppm{x}{\_}\begin{branch}
\mpom{a}{3}\ppm{a}{1}
\end{branch}\\\\
\mpom{x}{2}\ppm{a}{0}
\end{array}$

\medskip
An example of endless $\eta$-expansion is provided by the
well-known term $\Delta\Delta$  (with $\Delta=\lbd x.xx$):
$$(x(x))\Sub{x}{(\lbd y.y(y))}$$
The terms $(x(x))$ and $(\lbd y.y(y))$ are not $\eta$-long -- a concept which is
meaningless without types. So the machine has to make 
$\eta$-expansions dynamically.

\medskip
Strategy for $(x(x))$:

$$\comm{(}{\bullet}\comm{x_1}{\ppm{x}{\_}}\begin{branch}
\comm{(}{\pom{1}}\comm{x_2}{\ppm{x}{\_}}
\end{branch}$$

\medskip
Strategy for $\Sub{x}{(\lbd y.y(y))}$:

$$\comm{(\lbd y}{\pom{x}}\comm{y_1}{\ppm{1}{0}}\begin{branch}
\comm{(}{\pom{1}}\comm{y_2}{\ppm{1}{1}}
\end{branch}$$

We display the steps of the $\mbox{GAM}_\eta$ up to step $(\appst{21})$. Note the
(exponential) explosion of steps related with the traversal of the expansion variables that are progressively introduced. Notice also that the GAM is stuck as early as step $(\appst{5})$.

\medskip
Multiplexed, expanded strategy for
$(x(x))$:

\newcommand{\ppms}[1]{[\stackrel{#1}{\hookleftarrow}]} 
\newcommand{\mpoms}[1]{\langle 1,{\bf #1}\rangle} 

$$\mpom{\bullet}{1}\ppm{x}{\_}\begin{branch}
\mpoms{3}\ppm{x}{\_}\begin{branch}
\mpoms{5}\comm{x_2^{\eta}}{\ppms{1}}\begin{branch}
\mpoms{11}\comm{x_2^{2\eta}}{\ppms{1}}
\end{branch}\\
\mpoms{13}\comm{x_2^{\eta}}{\ppms{1}}\begin{branch}
\mpoms{19}\comm{x_2^{2\eta}}{\ppms{1}}
\end{branch}
\end{branch}\\
\mpoms{7}\ppm{x}{\_}\begin{branch}
\mpoms{9}\comm{x_2^{\eta}}{\ppms{1}}
\end{branch}\\
\mpoms{15}\ppm{x}{\_}\begin{branch}
\mpoms{17}\comm{x_2^{\eta}}{\ppms{1}}
\end{branch}
\end{branch}$$

\medskip
Multiplexed, expanded strategy for
$\Sub{x}{(\lbd y.y(y))}$:

$$\begin{branch}
\mpom{x}{2}\ppms{0}\begin{branch}
\mpoms{6}\ppms{1}\begin{branch}
\mpoms{10}\comm{y_2^{\eta}}{\ppms{1}}
\end{branch}\\
\mpoms{14}\ppms{1}\begin{branch}
\mpoms{18}\comm{y_2^{\eta}}{\ppms{1}}\end{branch}\\
\end{branch}\\
\mpom{x}{4}\ppms{0}\begin{branch}
\mpoms{12}\ppms{1}\begin{branch}
\mpoms{20}\comm{y_2^{\eta}}{\ppms{1}}
\end{branch}
\end{branch}\\
\mpom{x}{8}\ppms{0}\\
\mpom{x}{16}\ppms{0}
\end{branch}
$$

\subsection{B\"ohm's theorem} \label{xy-x(xy)}

B\"ohm's theorem asserts that in the $\lbd$-calculus one can separate two distinct normal forms $M_1$ and $M_2$ that are not $\eta$-convertible \cite{Bare}. This means that we can find a context  (that is, a term with a hole)
 $C$ such that $C[M_1]$ ($C$ whose hole has been filled with $M_1$) and $C[M_2]$ $\beta$-reduce
 to distinct (fresh) variables. This result in turn entails that $\beta\eta$ is a maximal consistent theory, since adding any new equation (between normalizable terms) would result in equating all pairs of terms.

As an illustration, we show  how to separate $xy$ et $x(xy)$. We follow
Joly's proof of B\"ohm's theorem \cite{JolyTh}.  Instead of distinct fresh variables, we use
distinct constants $\Omega$ and $\demon$, as in ludics (cf. section \ref{Examples-Ludics}).

Joly's proof makes use of auxiliary terms:

$$\begin{array}{l}
0=\lbd xy.y\\
1=\lbd xy.x\\
(M,N)_n=\lbd y_1\ldots y_nz.z(My_1\ldots y_n)(Ny_1\ldots y_n)\\
R=\lbd z.1\\
S=0
\end{array}$$

\noindent
We observe:
$$\begin{array}{l}
(M,N)_00\rightarrow N\\
(M,N)_01\rightarrow M\\
(M,N)_{k+1}P\rightarrow (MP,NP)_k
\end{array}$$

\noindent
The separating context is the following:
$$[\:][x\leftarrow(\lbd x.0,\lbd x.x)_3,y\leftarrow 1]RS01\Omega\demon$$

\noindent
We set $P\equiv y$ or $P\equiv xy$, and $P'\equiv P[x\leftarrow(\lbd x.0,\lbd x.x)_3,y\leftarrow 1]$. We have:
$$\begin{array}{l}
(xP)[x\leftarrow(\lbd x.0,\lbd x.x)_3,y\leftarrow 1]RS01\Omega\demon\\
\quad\equiv (\lbd x.0,(\lbd x.x)_3P'RS01\Omega\demon\\
\quad\rightarrow^* (\_,\lbd x.x)P'RS)_001\Omega\demon\\
\quad\rightarrow (\lbd x.x)P'RS1\Omega\demon\\
\quad\rightarrow P'RS1\Omega\demon
\end{array}$$
\begin{itemize}
\item $P=y$. Then $P'\equiv 1$ and:
$$\begin{array}{l}
P'RS1\Omega\demon \equiv 1RS1\Omega\demon\\
\quad\rightarrow R1\Omega\demon\equiv (\lbd z.1)1\Omega\demon\\
\quad\rightarrow 1\Omega\demon\\
\quad \rightarrow \Omega
\end{array}$$
\item $P=xy$. Then $P'\equiv (\lbd x.0,\lbd x.x)_31$ and:
$$\begin{array}{l}
P'RS1\Omega\demon \equiv (\lbd x.0,\lbd x.x)_31RS1\Omega\demon\\
\quad\rightarrow^* ((\lbd x.0)1RS,\_)_01\Omega\demon\\
\quad\rightarrow (\lbd x.0)1RS\Omega\demon\\
\quad\rightarrow  0RS\Omega\demon
\quad \rightarrow S\Omega\demon\\
\quad\rightarrow \demon
\end{array}$$

\end{itemize}
Here is the  execution (case $P\equiv y$). The compilation of
$$(zRS01\Omega\demon)[z\leftarrow (xy)[x\leftarrow Q,y\leftarrow 1]]$$
where
$$Q=(\lbd x_1x_2x_3z.z(ax_1x_2x_3)(bx_1x_2x_3))[a\leftarrow \lbd x.0,b\leftarrow \lbd x.x]$$
is as follows:

\medskip
$\begin{array}{ccc}
\bullet^0\ppm{z}{\_}^6\begin{branch}
1^3 \ppm{2}{0}^0\\
2^2 \ppm{2}{0}^0
\\
3^2  \ppm{2}{0}^0\\
4^2 \ppm{1}{0}^0
\\
5^0\ppm{\Omega}{\_}\\
6^0\ppm{\demon}{\_}
\end{branch} & &
z^0\ppm{x}{\_}^1
\begin{branch}
1^0 \ppm{y}{\_}^0
\end{branch}
\end{array}$

\medskip

\medskip
$x^4\ppm{4}{0}^2
\begin{branch}
1^0 \ppm{a}{\_}^3
\begin{branch}
1^0 \ppm{1}{2}^0
\\
2^0 \ppm{2}{2}^0 
\\
3^0 \ppm{3}{2}^0
\end{branch}
\\
2^0 \ppm{b}{\_}^3
\begin{branch}
1^0 \ppm{1}{2}^0
\\
2^0 \ppm{2}{2}^0 
\\
3^0 \ppm{3}{2}^0
\end{branch}
\end{branch}$

\medskip
$a^3 \ppm{3}{0}^0$

\medskip
$b^1 \ppm{1}{0}^0$

\medskip
$y^2  \ppm{1}{0}^0 $

\medskip\noindent And here is the dynamics:

\medskip
$\mpom{\bullet}{1}\ppm{z}{\_}\begin{branch}
\mpom{1}{17} \ppm{2}{0}\\
\mpom{3}{5}  \ppm{2}{0}\begin{branch}
\mpom{2}{29} \ppm{4}{1}
\end{branch}\\
\mpom{5}{33}\ppm{\Omega}{\_}
\end{branch}$

\medskip

$\mpom{z}{2} \ppm{x}{\_}
\begin{branch}
\mpom{1}{10} \ppm{y}{\_}
\begin{branch}
\mpom{1}{12} \ppm{1}{1}
\begin{branch}
\mpom{2}{22}  \ppm{2}{1}\\
\end{branch}
\\
\mpom{4}{24}  \ppm{4}{1}
\end{branch}
\\
\mpom{2}{16} \ppm{1}{1}
\begin{branch}
\mpom{2}{18} \ppm{2}{1}
\end{branch}
\\
\mpom{4}{4} \ppm{3}{1}
\begin{branch}
\mpom{2}{6} \ppm{2}{1}\begin{branch}
\mpom{2}{28} \ppm{2}{1}
\end{branch}
\\
\mpom{4}{30}  \ppm{4}{1}\end{branch}
\\
\mpom{6}{32}\ppm{5}{1}\end{branch}$

\medskip

\medskip

$\mpom{x}{3} \ppm{4}{0}
\begin{branch}
\mpom{2}{7} \ppm{b}{\_}
\begin{branch}
\mpom{1}{9} \ppm{1}{2}
\begin{branch}
\mpom{1}{13} \ppm{1}{1}
\begin{branch}
\mpom{2}{21} \ppm{2}{1}
\end{branch}
\\
\mpom{4}{25}  \ppm{4}{1}\\
\end{branch}
\\
\mpom{2}{15} \ppm{2}{2} \begin{branch}
\mpom{2}{19} \ppm{2}{1}
\end{branch}
\\
\mpom{5}{27} \ppm{2}{1}
\end{branch}
\\
\mpom{4}{31}  \ppm{6}{1}
\end{branch}$

\medskip

\medskip

$ \mpom{b}{8} \ppm{1}{0}
\begin{branch}
\mpom{1}{14} \ppm{2}{1}
\begin{branch}
\mpom{2}{20} \ppm{2}{1}
\end{branch}\\
\mpom{4}{26}  \ppm{5}{1}
\end{branch}$

\medskip

\medskip

$\mpom{y}{11}  \ppm{1}{0} \begin{branch}
\mpom{2}{23}\ppm{4}{1}
\end{branch}$

\bigskip \noindent Taking now $P=xy$, we replace the
tree starting with $z$ by the following one: 

\medskip

$z^0\ppm{x}{\_}^1
\begin{branch}
1^0 \ppm{x}{\_}^1\begin{branch}
1^0 \ppm{y}{\_}^0
\end{branch}
\end{branch}$

\medskip\noindent
The execution is the same until step 10, where the visit of $P$ starts. We give the full execution below:

\medskip
{\small $\mpom{\bullet}{1}\ppm{z}{\_}\begin{branch}
\mpom{2}{39} \ppm{2}{0}
\\
\mpom{3}{5}  \ppm{2}{0}\begin{branch}
\mpom{1}{17}\ppm{3}{1}\begin{branch}
\mpom{1}{25}\ppm{1}{1}\begin{branch}
\mpom{2}{53}\ppm{2}{1}
\end{branch}\\
\mpom{4}{61}\ppm{4}{1}

\end{branch}\\
\mpom{3}{73}\ppm{5}{1}

\end{branch}\\
\mpom{4}{21} \ppm{1}{0}\begin{branch}
\mpom{2}{57}\ppm{4}{1}
\end{branch}

\\
\mpom{6}{77}\ppm{\demon}{\_}
\end{branch}$

\medskip

$\mpom{z}{2}\ppm{x}{\_}
\begin{branch}
\mpom{1}{10} \ppm{x}{\_}\begin{branch}
\mpom{3}{34} \ppm{2}{1}\begin{branch}
\mpom{2}{44}\ppm{2}{1}
\end{branch}
\\
\mpom{4}{12} \ppm{3}{1}\begin{branch}
\mpom{1}{30}\ppm{1}{1}\begin{branch}
\mpom{2}{48}\ppm{2}{1}
\end{branch}\\
\mpom{4}{66}\ppm{4}{1}

\end{branch}\\
\mpom{6}{68}\ppm{5}{1}
\end{branch}\\
\mpom{3}{38}\ppm{2}{1}\begin{branch}
\mpom{2}{40}\ppm{2}{1}
\end{branch}\\
\mpom{4}{4}\ppm{3}{1}\begin{branch}
\mpom{2}{6}\ppm{2}{1}\begin{branch}
\mpom{1}{16}\ppm{1}{1}\begin{branch}
\mpom{1}{26}\ppm{1}{1}\begin{branch}
\mpom{2}{52}\ppm{2}{1}
\end{branch}\\
\mpom{4}{62}\ppm{4}{1}

\end{branch}\\
\mpom{3}{72}\ppm{3}{1}

\end{branch}\\
\mpom{3}{18}\ppm{3}{1}\begin{branch}
\mpom{1}{24}\ppm{1}{1}\begin{branch}
\mpom{2}{54}\ppm{2}{1}
\end{branch}\\
\mpom{4}{60}\ppm{4}{1}

\end{branch}\\
\mpom{5}{74}\ppm{5}{1}

\end{branch}
\\
\mpom{5}{20}\ppm{4}{1}\begin{branch}
\mpom{1}{22}\ppm{1}{1}\begin{branch}
\mpom{2}{56}\ppm{2}{1}
\end{branch}\\
\mpom{4}{58}\ppm{4}{1}

\end{branch}\\
\mpom{7}{76}\ppm{6}{1}
\end{branch}$

\medskip

\medskip

$\mpom{x}{3}\ppm{4}{0}
\begin{branch}
\mpom{2}{7} \ppm{b}{\_}
\begin{branch}
\mpom{1}{9} \ppm{1}{2}\begin{branch}
\mpom{2}{35}\ppm{2}{1}
\begin{branch}
\mpom{2}{43}\ppm{2}{1}
\end{branch}
\\
\mpom{3}{13}\ppm{3}{1}\begin{branch}
\mpom{1}{29}\ppm{1}{1}\begin{branch}
\mpom{2}{49}\ppm{2}{1}
\end{branch}\\
\mpom{4}{65}\ppm{4}{1}

\end{branch}\\
\mpom{5}{69}\ppm{5}{1}

\end{branch}
\\
\mpom{3}{37} \ppm{3}{2}\begin{branch}
\mpom{2}{41}\ppm{2}{1}
\end{branch}

\\
\mpom{4}{15}\ppm{1}{1}\begin{branch}
\mpom{1}{27}\ppm{1}{1}\begin{branch}
\mpom{2}{51}\ppm{2}{1}
\end{branch}\\
\mpom{4}{63}\ppm{4}{1}

\end{branch}\\
\mpom{6}{71}\ppm{3}{1}

\end{branch}
\\
\mpom{3}{19}\ppm{5}{1}\begin{branch}
\mpom{1}{23}\ppm{1}{1}\begin{branch}
\mpom{2}{55}\ppm{2}{1}
\end{branch}\\
\mpom{4}{59}\ppm{4}{1}

\end{branch}\\
\mpom{5}{75}\ppm{7}{1}
\end{branch}$

\medskip

$\mpom{x}{11}\ppm{4}{0}
\begin{branch}
\mpom{1}{31} \ppm{a}{\_}
\begin{branch}
\mpom{3}{33} \ppm{3}{2}\begin{branch}
\mpom{2}{45}\ppm{2}{1}
\end{branch}
\\
\mpom{5}{47}\ppm{2}{1}
\end{branch}
\\
\mpom{4}{67}\ppm{6}{1}
\end{branch}$

\medskip

$\mpom{a}{32} \ppm{3}{0}\begin{branch}
\mpom{2}{46}\ppm{5}{1}
\end{branch}
$

\medskip

$\mpom{b}{8} \ppm{1}{0}\begin{branch}
\mpom{2}{36}\ppm{3}{1}\begin{branch}
\mpom{2}{42}\ppm{2}{1}
\end{branch}
\\
\mpom{3}{14}\ppm{4}{1}\begin{branch}
\mpom{1}{28}\ppm{1}{1}\begin{branch}
\mpom{2}{50}\ppm{2}{1}
\end{branch}\\
\mpom{4}{64}\ppm{4}{1}

\end{branch}\\
\mpom{5}{70}\ppm{6}{1}

\end{branch}$}

\subsection{A degenerated general instance} \label{fax}

When there are neither arguments nor abstracted variables, then $\eta$-expansion makes sense
even for arbitrary B\"ohm trees: $(\cppm{a}{\star}\set{\:})$ expands to
$$(\lbd{\tt z}.\cppm{a}{\star}\setc{(a,\cppm{a}{1})}{a\in A'})$$
where $A'$ is an arbitrary subset of $A$. Taking $\cppm{a}{\star}=\cppm{x_i}{\_}$ and $A'=A$, and co-inductively iterating the process, we obtain the identity morphisms of our category, which completes the categorical picture of section \ref{Strong-section}.
We check on one example that this works, by playing $(\ppm{x_2}{\_}\set{\:})$ against

\medskip
$\begin{array}{l}
\begin{branch}
x_1\begin{branch}\ldots\end{branch}\\\\
x_2\ppm{a}{0} \begin{branch} b \ppm{y}{\_}\end{branch}\\\\
x_3\begin{branch} \ldots \end{branch}
\end{branch}
\end{array}$

\medskip\noindent
We $\eta$-expand $(\ppm{x_2}{\_}\set{\:})$ dynamically, using the (strong version of the) machinery as  in section \ref{GAM-eta}, replacing in the rules $m,n_1$, and $n_2$ by $a$ (arbitrary move), $0$  (empty set of abstracted variables), and $0$ (empty set of arguments), respectively:

\medskip
$\begin{array}{l}
\mpom{\bullet}{1} \ppm{x_2}{\_}\begin{branch}  \mpom{a}{3} \ppm{a}{1} \begin{branch}
\mpom{b}{4} \ppm{b}{1}
\end{branch}
\end{branch}\\\\\\
\begin{branch}
x_1\begin{branch}\ldots\end{branch}\\\\
\mpom{x_2}{2}\ppm{a}{0} \begin{branch} b \ppm{y}{\_}
\begin{branch}
\mpom{b}{5}Ê\ppm{y}{\_}
\end{branch}
\end{branch}\\\\
x_3\begin{branch} \ldots \end{branch}
\end{branch}
\end{array}$

\subsection{Discussion}  \label{Discussion-separation}
B\"ohm's  theorem works only modulo $\eta$-expansion (we refer to
\cite{CuEta} for a detailed analysis). But, as we have seen, $\eta$-conversion does not make sense
in the general setting of abstract B\"ohm trees. As a matter of fact, B\"ohm's theorem  does not hold \cite{DavPy} for the $\lbd\mu$-calculus in its original version, while only a small 
increase of flexibility in the syntax makes it
hold again  \cite{Saurin}.  In other words, B\"ohm's theorem is rather fragile and does not extend easily to other syntaxes.

\smallskip
What about separation in the strict sense (not modulo $\eta$)?
Maurel has exhibited the following simple counter-example to separation  in ludics with pointers:

$$\begin{array}{l}
M_1=(x\cdot \set{0})\set{\lbd\set{x_0}.\demon}\\
M_2=(x\cdot\set{0})\set{\lbd\set{x_0}.(x\cdot \set{0})\set{\lbd\set{x_0}.\demon}}
\end{array}$$

\noindent Here,
say, $\lbd\set{x_0}.\demon$ is a notation for
$$\set{\lbd\set{x_0}.\demon}\union\setc{J=\lbd\setc{x_j}{j\in J}.\Omega}{J\neq\set{0}}$$
 (This example is a variant of the terms $xy$ and $x(xy)$ considered in section \ref{xy-x(xy)}.)

\smallskip\noindent

The only (closed) opponents $N$ able to interact with $M_1$ in such a way that
$\coupe{M_1}{[x\leftarrow N]}$ converges have the form
$$\begin{array}{l}
\set{\ldots,(\set{0}=\lbd\set{y_0}.\demon),\ldots}\mbox{ or}\\
\set{\ldots,(\set{0}=\lbd\set{y_0}.(y_0\cdot\set{0})\set{M}),\ldots}
\end{array}$$

The opponents to $M_2$ must also have the same form. But then it is easily checked that for any $N$ in this class, the evaluations of {\it both} $\coupe{M_1}{[x\leftarrow N]}$ and 
$\coupe{M_2}{[x\leftarrow N]}$ converge. For example, with $N$ of the second form, we have
$$\begin{array}{lll}
\coupe{M_2}{[x\leftarrow N]} &
 \longrightarrow &
\coupe{y_0\cdot\set{0})\set{M}}{[y_0\leftarrow M_1[x\leftarrow N]]}\\
& \longrightarrow & \coupe{M_1}{[x\leftarrow N]}\\
& \longrightarrow & \coupe{y_0\cdot\set{0})\set{M}}{[y_0\leftarrow
(\lbd\set{x_0}.\demon)[x\leftarrow N]]}\\
& \longrightarrow & \coupe{\demon}{\ldots}
\end{array}$$

But why have we been able to separate $xy$ and $x(xy)$?
We can reformulate the separating context as follows. We have let 
$$\begin{array}{l}
M_1=\lbd xy.xy\quad\mbox{ and}\quad
M_2=\lbd xy.x(xy)
\end{array}$$
interact with $zQ1RS01\Omega\demon$, through the substitution of $M_1$ or $M_2$ for $z$.
But in terms of ludics, this means that we have accepted an interaction 
between $(z\cdot\set{1,2,3,4,5,6,7,8}{\ldots}$ and $\lbd\set{1,2}.\ldots$, which violates the
machinery of ludics (and of abstract B\"ohm trees in general). The point of $\eta$-expansion is
to allow to fill the gap, and to pretend that both $M_1$ and $M_2$ have arity 8.

\medskip Separation is recovered under Girard's affinity conditions, and the proof is then simple (see \cite{GirLS,CuLL-ludintro}). In this respect, the situation is the same  as for B\"ohm's theorem, which  becomes tricky only when a head variable occurs in one of its arguments (like in $y(yx)$). 

Maurel has shown how to recover separation without sacrificing nested occurrences of variables  by extending the framework of designs to probabilistic designs \cite{MaurelTh}.  The idea is to assign probabilities to actions, i.e., to head variables. When assigning
probability $\frac{1}{2}$ to $y_0\cdot\set{0}$ in  $\lbd\set{y_0}.(y_0\cdot\set{0})\set{M}),\ldots$,
we get that $\coupe{M_1}{[x\leftarrow N]}$ and $\coupe{M_2}{[x\leftarrow N]}$ converge with probabilities $\frac{1}{2}$ and $\frac{1}{4}$, respectively, and hence $M_1$ and $M_2$ can be separated. 

\smallskip
We summarize the discussion in figure \ref{Separation-YN}.

\begin{figure}
\Figbar

\begin{center}
\begin{tabular}{ccc}
& exponential ludics : NO &\\\\
(affine) ludics: YES 	&& exponential probabilistic ludics : YES\\\\
& $\lbd$-calculus: yes (modulo $\eta$) &
\end{tabular}
\end{center}
\caption{Three separation theorems}
\label{Separation-YN}
\Figbar
\end{figure}

\medskip
We believe that Maurel's probabilistic ludics can be lifted to the general setting of abstract B\"ohm trees, and leave this as further work.

\appendix
\section{Machine equivalences}
We define precise translations between the machines. The translation functions are written $\trad{\_}{XY}$ (from machine $X$ to machine $Y$).

\subsection{VAM-GAM equivalence} \label{VAM-GAM}
We define two-way translations between the GAM and the VAM.
The GAM state associated to a VAM state is (essentially) its associated set of multiplexed views, which are defined as follows:

$$\begin{array}{l}
\seq{}
{\dview{\Gamma}{1}=\bullet}\quad\quad\seq{\position{\Gamma}{(2n)}=a}
{\dview{\Gamma}{2n}=\mpom{a}{2n}}\\\\
\seq{\position{\Gamma}{n}=\jitem{a}{\appst{m}}}
{\dview{\Gamma}{n}=\dview{\Gamma}{m-
1}\:(\position{\Gamma}{\appst{m}})\:\mpom{a}{n}}
\end{array}$$

\smallskip\noindent
VAM to GAM:
$$\begin{array}{c}
\seq{}{\trad{\litem{\bullet}{1}}{VG}=\litem{\mpom{\bullet}{1}}{1}} 
\\\\
\seq{}{\trad{\Gamma\litem{\_}{n}}{VG}=\trad{\Gamma}{VG}\litem{\dview{\Gamma}{n}}{n}} \\\\
\seq{\position{\trad{\Gamma}{VG}}{(n-1)}=\dyn{q}}
{\trad{\Gamma\litem{\ppm{a}{\kappa}}{\appst{n}}}{VG}=\trad{\Gamma}{VG}\litem{\dyn{q}\ppm{a}{\kappa}}{\appst{n}}}
\end{array}$$
where $\_$ stands for either $a$ or $\vitem{a}{\appst{m}}$.

\medskip
The GAM to VAM direction is essentially forgetful:

\smallskip\noindent
GAM to VAM:
$$\begin{array}{c}
\seq{}{\trad{\litem{\mpom{\bullet}{1}}{1}}{GV}=\litem{\bullet}{1}} \\\\
\seq{}{\trad{\Gamma\litem{\dyn{q}\ppm{a}{\kappa}}{\appst{n}}}{GV}=\trad{\Gamma}{GV}\litem{\ppm{a}{\kappa}}{\appst{n}}} \\\\
\seq{\position{\Gamma}{\appst{2n}}=\dyn{q}\ppm{a}{\_}}{\trad{\Gamma\litem{\mpom{a}{2n}}{2n}}{GV}=
 \trad{\Gamma}{GV}\litem{a}{2n}} \\\\
\seq{\pi'(\pop^i(\position{\Gamma}{(n-1)}))=m}
{\trad{\Gamma\litem{\dyn{r}\mpom{a}{n}}{n}}{GV} = \trad{\Gamma}{GV}\litem{\vitem{a}{\appst{m}}}{n}}
\end{array}$$

\noindent
We shall show that these transformations are inverse (on reachable states, i.e., on states that arise at some stage in the execution of the machine), and that the two machines simulate each other in lock step. We shall use the following invariants:
$$\begin{array}{ll}
(1) & \pi'(\pop^i(\dview{\Gamma}{n}))=\jump{\Gamma}{i}{n} \quad (n>1)\\
(2) & \dview{\trad{\Gamma}{GV}}{n}=\position{\Gamma}{n} \quad\mbox{(for $\Gamma$ reachable)}
\end{array}$$
We prove claim (1) by induction on $i$:
\begin{enumerate}
\item $i=0$. Then the claim reduces to $ \pi'(\dview{\Gamma}{n})=n$, which holds by definition of the multiplexed view function.
\item $i>0$. Then $\position{\Gamma}{n}=\vitem{a}{\appst{m}}$ for some $\appst{m}$. We have, by definition of {\it jump} and {\it dvview}:
$$m-1=\jump{\Gamma}{}{n}\quad\mbox{and}\quad \dview{\Gamma}{n}=\dview{\Gamma}{m-
1}\:(\position{\Gamma}{\appst{m}})\:\mpom{a}{n}$$
It follows that the claim reduces to its $(i-1,m-1)$ instance, which holds by induction.
\end{enumerate} 
The base cases of claim (2) are obvious. For the induction case, we have
$$\begin{array}{lll}
\dview{\trad{\Gamma}{GV}}{n} & = & \dview{\trad{\Gamma}{GV}}{m-
1}\:(\position{\trad{\Gamma}{GV}}{\appst{m}})\:\mpom{a}{n}\\
&=&  (\position{\Gamma}{(m-1)})\:(\position{\trad{\Gamma}{GV}}{\appst{m}})\:\mpom{a}{n} \\
&=& (\position{\Gamma}{\appst{m}})\:\mpom{a}{n}\\
&=& \position{\Gamma}{n}
\end{array}$$
using successively the induction hypothesis,  the definition of the $\appst{m}$-rule of the GAM, the definition of the translation, and the definition of the $n$-rule of the GAM.  Appeals to the rules of the GAM are legitimate because the claim is restricted to reachables states.


\smallskip We now show that the machines simulate each other in lock-step.
Let $\litem{\mpom{\bullet}{1}}{1}=\Gamma_1,\ldots,\Gamma_\nu,\ldots$ be the successive states of the GAM.
Then the VAM runs successively  through the states $\litem{\bullet}{1}=\trad{\Gamma_1}{GV},\ldots\trad{\Gamma_\nu}{GV},\ldots$, and terminates only when the GAM terminates; and the same holds conversely in the VAM to GAM direction.
These properties are an easy consequence of the two claims. We check one case in detail:
let $\Gamma\mapsto\Gamma\litem{\dyn{r'}\mpom{a}{2n}}{2n}$, with $\pi'(\pop^i(\position{\Gamma}{(2n-1)}))=2m-1$.  Using the two claims, we have:
$$\pi'(\pop^i(\position{\Gamma}{(2n-1)})) = \pi'(\pop^i(
 \dview{\trad{\Gamma}{GV}}{2n-1})) = \jump{\trad{\Gamma}{GV}}{i}{2n-1}$$
 from which it follows that 
 $$\trad{\Gamma}{GV}\mapsto\trad{\Gamma}{GV}\litem{\vitem{a}{\appst{2m-1}}}{2n}=\trad{\Gamma\litem{\dyn{r'}\mpom{a}{2n}}{2n}}{GV}$$
 The fact that the two translations are inverse  on reachable states is an immediate consequence of the mutual lock-step simulation.

\subsection{SAM-GAM equivalence}
 We follow the same scheme as in section \ref{VAM-GAM}.
Since the proofs are fairly similar, we shall limit ourselves to stating
the relevant invariants.

 If $\Gamma$ is a state of the GAM, we use $\Gamma_\nu$ to denote the prefix of $\Gamma$ whose last item is the $\nu$-th item of $\Gamma$. 
The inverse translations are defined as follows (the second translation necessitates to define
at the same time a step number associated to a SAM state):

\smallskip\noindent
GAM to SAM:
$$\begin{array}{c}
\seq{}
{\trad{\litem{\mpom{\bullet}{1}}{1}}{GS}=\bullet}\\\\
\seq{\hd{\Gamma}=\litem{\dyn{q}\ppm{a}{i}}{\appst{n}} \quad \dnumber{\pop^i(\dyn{q})}=m}
{\trad{\Gamma\litem{\dyn{r'}\mpom{a}{n}}{n}}{GS}=\samenv{\appst{m}}\vitem{a}{\trad{\Gamma}{GS}}} \\\\
\seq{\hd{\Gamma}=\litem{\dyn{q}\ppm{a}{\_}}{\appst{2n}}}
{\trad{\Gamma\litem{\mpom{a}{2n}}{2n}}{GS}=\vitem{a}{\trad{\Gamma}{GS}}} \\\\
\seq{}{\trad{\Gamma\litem{\dyn{q}\ppm{a}{\kappa}}{\appst{n+1}}}{GS}=\trad{\Gamma}{GS}\ppm{a}{\kappa}}
\end{array}$$

\smallskip\noindent
SAM to GAM:
$$\begin{array}{ccc}
\seq{}{\trad{\bullet}{SG}=\litem{\mpom{\bullet}{1}}{1}} && \seq{}{\bullet^\#=1}\\\\
\seq{{\tt q}^\#=n}{\trad{{\tt q}\ppm{a}{\kappa}}{SG}=\trad{{\tt q}}{SG}\litem{(\position{\trad{{\tt q}}{SG}}{n})\ppm{a}{\kappa}}{\appst{n+1}}} &&
\seq{{\tt q}^\#=n}{({\tt q}\ppm{a}{\kappa})^\#=\appst{n+1}}\\\\
\seq{{\tt r}^\#=\appst{n}\quad {\tt r'}^\#=\appst{m}}
{\trad{{\tt r'}\vitem{a}{{\tt r}}}{SG}=\trad{{\tt r}}{SG}\litem{(\position{\trad{{\tt r'}}{SG}}{\appst{m}})\mpom{a}{n}}{n}} &&
\seq{{\tt r}^\#=\appst{n}}{({\tt r'}\vitem{a}{{\tt r}})^\#=n}\\\\
\seq{{\tt r}^\#=\appst{2n}}
{\trad{\vitem{a}{{\tt r}}}{SG}=\trad{{\tt r}}{SG}\litem{\mpom{a}{2n}}{2n}} &&
\seq{{\tt r}^\#=\appst{2n}}{\vitem{a}{{\tt r}}^\#=2n}
\end{array}$$

\noindent In order to express one of the invariants of these translations, we introduce the following substate relation among SAM states. It is the transitive closure of the relation defined by the following rules:
$$\begin{array}{lllll}\seq{}{{\tt q}\prec{\tt q}\ppm{a}{\kappa}} &&
\seq{}{{\tt r'}\prec{\tt r'}\vitem{a}{{\tt r}}} &&
\seq{}{{\tt r}\prec{\tt r'}\vitem{a}{{\tt r}}}
\end{array}$$
We also need the following auxiliary definition.
Let $\sigma$ be the function assigning to $\nu$ the state reached by the SAM at stage $(\nu)$, i.e.,
$\hfs{1} \bullet=\sigma(1) \hfs{\appst{2}} \sigma(\appst{2}) \ldots \hfs{\nu}\sigma(\nu) \ldots$
We are now ready to express the invariants needed to prove the equivalence between the GAM and 
the SAM:

$$\begin{array}{l}
(\sigma(\nu))^\#=\nu \\
\sigma({\tt p}^\#)={\tt p}\\
{\tt p}_1\prec{\tt p}_2$, then $({\tt p}_1)^\#<({\tt p}_2)^\#\\
\mbox{the head item of }\trad{{\tt p}}{SG}\mbox{ is numbered } {\tt q}^\#\\
(\trad{\Gamma_\nu}{GS})^\#=\nu\\
(\trad{{\tt p}_1}{SG})_{({\tt p}_2)^\#}=\trad{{\tt p}_2}{SG} \quad ({\tt p}_2\prec{\tt p}_1)\\
(\pi'(\pop^i(\position{\trad{{\tt q}}{SG}}{{\tt q}^\#})))'=(\pi'(\pop^i({\tt q})))^\#\\
\samenv{(\dnumber{\pop^i(\position{\Gamma}{n})})'}=\dnumber{\pop^i(\samenv{n})}\\
\erase{\samenv{n}}=\erase{\position{\Gamma}{n}}
\end{array}$$

\subsection{SAM-EAM equivalence}
We relate the SAM to the EAM.  Let 
$$\coupe{P}{\cppm{a_1}{\_}\leftarrow M_1[\:],\ldots
\cppm{a_n}{\_} \leftarrow M_n[\:]}$$
be an initial state of the EAM. We shall show that the  EAM execution from there is lock-step simulated by the SAM execution of
$$\coupe{(\trad{P}{ca}_{[\:]})}{a_1\leftarrow \trad{M_1}{ca}_{[\:]},\ldots,a_1\leftarrow \trad{M_n}{ca}_{[\:]}}$$

\noindent
We translate a SAM state back to an EAM state as follows:

$$\begin{array}{l}
\seq{\begin{array}{l}
\trad{{\tt r}'}{SE}=\coupe{\ppm{a_1}{\star}\set{\ldots,(a,(\lbd{\tt y}.P)),\ldots}}{\rho'}\\
\trad{{\tt r}}{SE}=\coupe{\ppm{b_1}{\star}\setc{(b,M_b)}{b\in B}}{\rho}
\end{array}}
{\trad{{\tt r}'\vitem{a}{{\tt r}}}{SE}=\coupe{P}{\rho'[\cppm{b}{{\tt y}}\leftarrow M_b\mid b\in B]}}
\end{array}$$
$$\begin{array}{l}
\seq{\begin{array}{l}
\trad{{\tt r}'}{SE}=\coupe{\ppm{a_1}{\star}\set{\ldots,(a,(P)),\ldots}}{\rho'}\\
\trad{{\tt r}}{SE}=\coupe{\ppm{b_1}{\star}\set{\:}}{\rho}
\end{array}}
{\trad{{\tt r}'\vitem{a}{{\tt r}}}{SE}=\coupe{P}{\rho'}}
\end{array}$$

\noindent
In order to formulate the invariants of the simulation, we need to define the following translation function:

$$\begin{array}{l}
\seq{}{\trad{\bullet}{ac}=((P),[\:])}\quad\quad\quad\quad
\seq{}{\trad{a_i}{ac}=(M_i,[\:])}\\\\
\seq{\trad{r}{ac}=(\cppm{a}{\star}\set{\ldots,(b,M_b),\ldots},L)}
{\trad{rb}{ac}=(M_b,L)}\\\\
\seq{\trad{q}{ac}=((\lbd{\tt x}.P),L)}
{\trad{q\ppm{a}{\kappa}}{ac}=(P,{\tt x}\cons L)}
\end{array}$$

\noindent
The function $\trad{\_}{ac}$ reconstructs the matching subterm of $(P)$ or of the $M_i$, together with the concrete binding information above it. The translation $\trad{\_}{SE}$ satisfies the following invariants:

$$\begin{array}{l}
\seql{(1)}{\trad{{\tt r}}{SE}=\coupe{P}{\rho}}{P=\trad{\erase{r}}{ac}}
\\\\
\seql{(2)}{\trad{{\tt q}}{SE}=\coupe{P}{\rho}\quad\quad
\trad{\pi'(\pop^i({\tt q}))}{SE}=\cppm{a_1}{\star}\set{\ldots,(a,M),\ldots}}
{\rho(\cppm{a}{\pop^i(\trad{\erase{{\tt q}}}{ac}})=M}
\end{array}$$

\noindent which in turn entail  (easy check) that the SAM can proceed  from ${\tt r}$ (to  ${\tt q}$) if and only if the EAM can proceed from $\trad{{\tt r}}{SE}$ (to $\trad{{\tt q}}{SE}$).

\begin{figure}
\Figbar
$$
 \xymatrix @C=0.4cm  {&&& GAM& \\
 EAM \ar @{-}[rr] && SAM  \ar @{-}[ru] && VAM  \ar @{-}[lu] }
$$
 \caption{A unique computing device in four disguises}
 \label{four-machines}
\Figbar
 \end{figure}
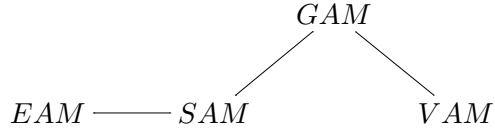

\medskip
Our picture is now complete. We have four equivalent presentations of the
same computing device, as summarized in figure \ref{four-machines}.

\end{document}